\documentclass[a4paper,11pt]{article}
\pdfoutput=1 
\usepackage{jcappub} 
\usepackage[T1]{fontenc} 

\newcommand{\bm}[1]{\hbox{\boldmath{$#1$}}}
\newcommand{\sbm}[1]{\hbox{\boldmath{\scriptsize$#1$}}}

\title{\boldmath Can axion clumps be formed in a pre-inflationary scenario?}

\author[a]{Hayato Fukunaga,}
\author[b,c]{Naoya Kitajima,}
\author[a,d]{Yuko Urakawa}
\affiliation[a]{Department of Physics and Astrophysics, Nagoya University, Chikusa, Nagoya 464-8602, Japan}
\affiliation[b]{Frontier Research Institute for Interdisciplinary Sciences, Tohoku University, Sendai, 980-8578 Japan}
\affiliation[c]{Department of Physics, Tohoku University, Sendai, 980-8578 Japan}
\affiliation[d]{Fakult{\"u}t f{\"u}r Physik, Universit{\"a}t Bielefeld, Bielefeld, 33501, Germany}

\emailAdd{fukunaga.hayato@c.mbox.nagoya-u.ac.jp}
\emailAdd{kitajima@tuhep.phys.tohoku.ac.jp}
\emailAdd{urakawa.yuko@h.mbox.nagoya-u.ac.jp}

\abstract{The QCD axion and an axion-like particle (ALP) are compelling candidates of dark matter. For the QCD axion, it is known that when the Peccei-Quinn (PQ) symmetry is spontaneously broken after inflation, the large initial fluctuation can lead to axion clump formation. On the other hand, when the symmetry is already broken during inflation, it has been believed that the axion clump formation does not occur due to the small amplitude of the initial axion fluctuation. We revisit this prevailing understanding, considering both the QCD axion and an ALP. We find that for the QCD axion, the clump formation does not occur even if we consider an extremely fine-tuned initial condition. Meanwhile, it turns out that for an ALP which allows a more general potential form, the clump formation can take place through the tachyonic instability or/and the resonance instability, considering a multiple cosine potential.}

\begin{document}
\maketitle
\flushbottom

\section{Introduction}
\label{sec:intro}
Dark matter (DM) is one of the puzzles in cosmology and particle physics to be solved. Since this essence cannot be explained within the standard model of particle physics, a new physics beyond the standard model is required to approach DM. Up until now, a variety of DM candidates have been considered. Axion or more generally axion-like particle (ALP) is one of the best-motivated candidates of DM.

Axion is a hypothetical particle introduced to solve the strong CP problem in quantum chromodynamics (QCD). The strong CP problem is related to a term that breaks CP symmetry in the QCD Lagrangian. The CP violation is characterized by a constant parameter $\theta_{{\rm QCD}}$, which can be examined by measuring the neutron electric dipole moment. From the current measurement, $\theta_{{\rm QCD}}$ is bounded as $|\theta_{{\rm QCD}}|\lesssim 10^{-10}$~\cite{Graner:2016ses}. Such an unnaturally small CP violation cannot be explained within the standard model. In the Peccei-Quinn (PQ) mechanism, which was proposed to solve this problem~\cite{Peccei:1977hh,Peccei:1977ur}, the $\theta_{\rm QCD}$ term goes to zero dynamically. The additionally introduced field for a realization of this mechanism is called QCD axion~\cite{Weinberg:1977ma,Wilczek:1977pj}. Soon after the original QCD axion model was ruled out because of the inconsistency with the electroweak physics, the invisible axion models were proposed \cite{Kim:1979if,Shifman:1978bx, Dine:1981rt, Zhitnitsky:1980}. In Refs.~\cite{Preskill:1982cy, Abbott:1982af,Dine:1982ah}, it was pointed out that the invisible axion can play the role of DM, when it oscillates coherently around the potential minimum. For a review, see e.g., Refs.~\cite{Sikivie:2006ni, Marsh:2015xka, Irastorza:2018dyq}.

Regarding cosmological aspect of the QCD axion, the key indicator is whether the PQ symmetry was already broken during inflation or has been broken after inflation. The former is the case when the PQ symmetry breaking scale, $f$, is comparable to or larger than the energy scale of inflation, characterized by the Gibbons-Hawking temperature, $T_{\rm GH} = H_I/2\pi$, with $H_I$ being the Hubble parameter during inflation, while the latter is the case when $f$ is smaller than $T_{\rm GH}$. The former is called pre-inflationary scenario and the latter is called post-inflationary scenario. When the PQ symmetry is already broken during inflation, satisfying $f > H_I/2\pi$, the inflation makes the axion field value almost homogeneous at least in our observable patch of the Universe. Then, the field fluctuation, corresponding to isocurvature fluctuation, is typically small, being consistent with CMB observations. On the other hand, when the PQ symmetry is restored during inflation, i.e. $f<H_I/2\pi$, the spontaneous symmetry breaking occurs after inflation, which results in a lot of causally-disconnected patches that have different initial field values. It leads to the formation of topological defects \cite{Sikivie:1982qv}. After the decay of topological defects, the field fluctuation can be $\mathcal{O}(1)$ on sub-horizon scales.

In the post-inflationary scenario, Kolb and Tkachev \cite{Kolb:1993zz, Kolb:1993hw, Kolb:1994fi} pointed out that when the axion potential switches on in the QCD epoch and the the axion commences the coherent oscillation, the fluctuation in the initial misalignment is transformed into the fluctuation in the energy density of the axion. Subsequently, the non-linear dynamics driven by the attractive self-interaction of the axion leads to formation of overdense axion clumps, called axiton~\cite{Kolb:1993hw} or axion miniclusters~\cite{Hogan:1988mp}. The formed axion overdense clumps undergo the non-linear gravitational collapse, as was addressed by conducting the $N$-body simulation in Ref.~\cite{Zurek:2006sy}. More recently, in Ref.~\cite{Eggemeier:2019khm}, the formation and clustering of axion minihalos were studied based on $N$-body simulation to date. In Ref.~\cite{Tinyakov:2015cgg}, it was argued that the tidal disruption may have a significant impact on the succeeding evolution. A semi-analytic computation of the mass function for axion miniclusters can be found in Ref.~\cite{Enander:2017ogx}. A review on axion miniclusters can be found e.g., in  Ref.~\cite{Tkachev:2015usb}.

The formation of axion clumps can leave an interesting phenomenological consequence, opening a window for axion search. In Refs.~\cite{Kolb:1993zz, Seidel:1993zk}, it was conjectured that axion stars, which are gravitationally bound and stable, may appear in the center of axion miniclusters through further contraction. More recent studies on axion stars include \cite{Braaten:2015eeu, Levkov:2016rkk, Visinelli:2017ooc, Niemeyer:2019aqm}. Axion clumps can cause gravitational lensing effect, which may be detectable through femtolensing and picolensing measurements~\cite{Kolb:1995bu} and microlensing measurements \cite{Fairbairn:2017dmf,Fairbairn:2017sil,Dai:2019lud}.

On the other hand, the possibility of the clump formation has not been well investigated, when the PQ symmetry was already broken during inflation, i.e., in pre-inflationary scenario. This is presumably because the initial fluctuation of the axion has been considered to be too small for the formation of axiton or axion miniclusters. Meanwhile, especially in the context of reheating, it is known that a similar overdense clump, called oscillons \cite{Gleiser_1994,Salmi:2012ta,Copeland_1995,Amin_2010,Amin_2012,Lozanov_2018,Olle:2019kbo}, can be formed for a wide class of scalar field models with an attractive self-interaction, even if we start with an almost homogeneous initial condition. As we increase the misalignment angle, being away from the potential minimum, the quartic term, which yields an attractive force, becomes more important. In an extreme case where the axion was located around the hilltop of the potential just after it acquires the potential, the fluctuation of the axion is expected to be enhanced by the tachyonic instability. Furthermore, since the tachyonic mass of the axion, $- m^2$, increases with time through the QCD instanton effect, one may expect that the tachyonic instability can be more efficient in the QCD epoch. A possible impact of the tachyonic instability on the axion isocurvature was studied in Ref.~\cite{Kobayashi:2013nva} by taking the large scale limit, where the $\delta N$ formalism~\cite{Starobinsky:1982ee, Sasaki:1995aw, Sasaki:1998ug} can be adopted. In this paper, we study the possibility of QCD axion clump formation in the pre-inflationary scenario, while accepting a fine-tuned initial condition. In Ref.~\cite{Co:2018mho}, a dynamical mechanism which places the axion at the hilltop was proposed. For our purpose, we should also take into account the spatial gradient of the axion configuration, in contrast to the situation addressed in Ref.~\cite{Kobayashi:2013nva}.

The compelling aspect of the QCD axion which enables to play the role of dark matter motivates us to consider a more general framework of axion like particle (ALP). Furthermore, ALPs are ubiquitously predicted in low energy effective field theories of string theory \cite{Witten:1984dg} (see also Refs.~\cite{Svrcek:2006yi, Arvanitaki:2009fg}). ALPs have various common aspects with the QCD axion e.g., behaving as non-relativistic matter during the coherent oscillation and an anomalous coupling with photons. Meanwhile, unlike the QCD axion, the shape and height of ALP potential are not necessarily determined by the QCD dynamics. This leaves a wider parameter space for ALPs to become dark matter. Taking the uncertainty of the ALP potential granted, in this paper, we also address the possibility of ALP clump formation for a generalized form of the potential that includes several cosine terms. We maintain the quadratic form around the potential minimum so that the ALP can become dark matter at a later time. Such a multiple cosine potential has been discussed for axion inflation models~\cite{Czerny:2014wza,Czerny:2014xja}.

The ALP potential with two cosine terms can have a shallower region than the quadratic form. It has been widely known that the tachyonic instability and/or the resonance instability can be significant for such a potential (see e.g., Refs.~\cite{Greene:1998pb,Soda:2017dsu,Fukunaga:2019unq}). When the potential is shallow enough, subsequently, overdense clumps, called oscillon, can be formed. It was pointed out that during the oscillon formation process, the gravitational waves can be copiously produced by considering axion inflation \cite{Zhou:2013tsa,Antusch:2016con,Liu:2017hua,Antusch:2017vga,Antusch:2017flz} and axion dark matter \cite{Soda:2017dsu, Kitajima:2018zco}. In Refs.~\cite{Adshead:2016iae, Patel:2019isj}, it was shown that these instabilities of ALPs also can provide a primordial origin of intergalactic magnetic fields.

This paper is organized as follows. In Sec.~\ref{sec:property} we briefly summarize the property of QCD axion and ALP. In Sec.~\ref{sec:clump_QCD}, we investigate the clump formation of the QCD axion when the PQ symmetry was already broken during inflation. In Sec.~\ref{sec:clump_ALPs}, we analyze clump formation for ALPs, clarifying several different formation processes. Finally, in Sec.~\ref{sec:conc}, we summarize our results.

\section{Basic property of QCD axion and ALPs}
\label{sec:property}
In this section, first we describe the setup of the problem with a brief review. Focusing on the axion which commences its oscillation during radiation domination, we ignore the contribution of the axion to the geometry of the Universe.

\subsection{Axion potential and relic abundance}
At the QCD phase transition, the QCD axion acquires the potential through the QCD instanton effect. The height of the potential is determined by the QCD scale, $\chi_{{\rm QCD}}$, which is called the QCD topological susceptibility. Here we adopt the results of the lattice QCD simulation in Ref.~\cite{Borsanyi:2016ksw}, where $\chi_{{\rm QCD}}$ is approximately given by
\begin{equation}
    \chi_{{\rm QCD}}(T)=\frac{\chi_0}{1+(T/T_c)^b}, \label{eq:chiQCD}
\end{equation}
with $\chi_0=(7.6\times10^{-2}\ {\rm GeV})^4$, $T_c=0.16$ GeV and $b=8.2$. Here, $T$ should be understood as the temperature of the relativistic species, which dominate the Universe. When the dilute instanton gas approximation (DIGA) holds, the potential of the QCD axion, $\phi$, is given by 
\begin{equation}
    V(\phi)=\chi_{{\rm QCD}}(T)(1-\cos(\phi/f)).
    \label{eq:potcos}
\end{equation}
The potential form slightly differs from the one computed from the low energy effective Lagrangian in Ref.~\cite{DiVecchia:1980yfw}. A detailed study about the potential form of the QCD axion and the coupling with the standard model sector can be found, e.g., in Refs.~\cite{diCortona:2015ldu, Fox:2004kb, Berkowitz:2015aua, Borsanyi:2015cka}.

For $T > T_{\rm c}$, as the temperature decreases, the mass of axion, given by
\begin{equation}
    m(T)= \sqrt{\chi_{\rm QCD}(T)}/f\,,
\end{equation}
increases as $m(T)\propto T^{-b/2}$. Meanwhile, for $T < T_{\rm c}$, $m(T)$ approaches to a constant value, given by
\begin{equation}
m_0 \simeq 6\,\mu{\rm eV}\left(\frac{10^{12}\,{\rm GeV}}{f} \right).
\end{equation} 
The axion commences oscillation at $T_{\ast}$ satisfying $3H(T_\ast) = m(T_\ast) \equiv m_\ast$ with $H(T)$ the Hubble parameter, and for $T_\ast > T_c$, $T_\ast$ is given by
\begin{equation}
T_\ast \simeq 1.0\,{\rm GeV} \left(\frac{10^{12}\,{\rm GeV}}{f}\right)^{0.16}.
\end{equation}
Then, one can obtain the final QCD axion abundance as (see also Refs.~\cite{Turner:1985si, Fox:2004kb,Visinelli:2009zm})
\begin{equation}
\Omega_a h^2 = 0.10 \kappa \left(\frac{f\tilde\phi_i}{10^{12}\,{\rm GeV}} \right)^{1.16} F(\tilde\phi_i),
\end{equation}
where $\kappa$ is ${\cal O}(1)$ numerical factor characterizing nonadiabaticity and $F(\tilde\phi_i)$ is anharmonic correction factor which is a function of the initial value of $\tilde\phi = \phi/f$, which becomes significant when the initial value is placed near the top of the cosine potential. The anharmonic correction $F(\tilde\phi_i)$ was computed, e.g., in Refs.~\cite{Turner:1985si, Lyth:1991ub, Bae:2008ue}. Note that the QCD axion with the decay constant being $f \sim 10^{12}$\,GeV and the initial value being $\tilde\phi_i \sim 1$ saturates the DM relic abundance in the Universe, corresponding to the upper bound of the so called "classical axion window"~\cite{ABBOTT1983133}, given by  
\begin{equation}
    4\times10^{8}{\rm GeV}<f<10^{12}{\rm GeV},  \label{Eq:rangef}
\end{equation}
where the lower bound comes from neutrino burst duration of SN1987A~\cite{Mayle:1987as}. Note that this upper bound includes a large ambiguity, because $\Omega_ah^2$ highly depends on the cosmological scenario.

\subsection{Potential of ALPs}\label{sec:alignedpotential}
In the case of ALPs, the potential can take a more general form, since it is not necessarily determined by the QCD physics. It is known that ALPs predicted in string theory generically acquire multiple cosine terms through the non-perturbative corrections as 
\begin{equation}
    V(\phi)=-\sum_{i=1}^n \Lambda_i^4\cos\left(\frac{\phi}{f_i}+\Theta_i\right)+{\rm const.},
\end{equation}
where $\Lambda_i$, $f_i$ and $\Theta_i$ are model parameters. An inflation model with multiple cosine terms was studied e.g., in alignment models~\cite{Kim:2004rp, Peloso:2015dsa} and multi-natural inflation model~\cite{Czerny:2014wza}.

% In particular, one can consider more general shape for the potential other than the single cosine type, such as those adopted by the pure-natural \cite{Nomura:2017ehb} or multi-natural \cite{Czerny:2014wza} inflation models.
% In the multi-natural inflation model, the potential is formed by multiple sinusoidal functions each of which has independent height, period and phase shift,

\begin{figure}
    \centering
    \includegraphics[height=5.5cm]{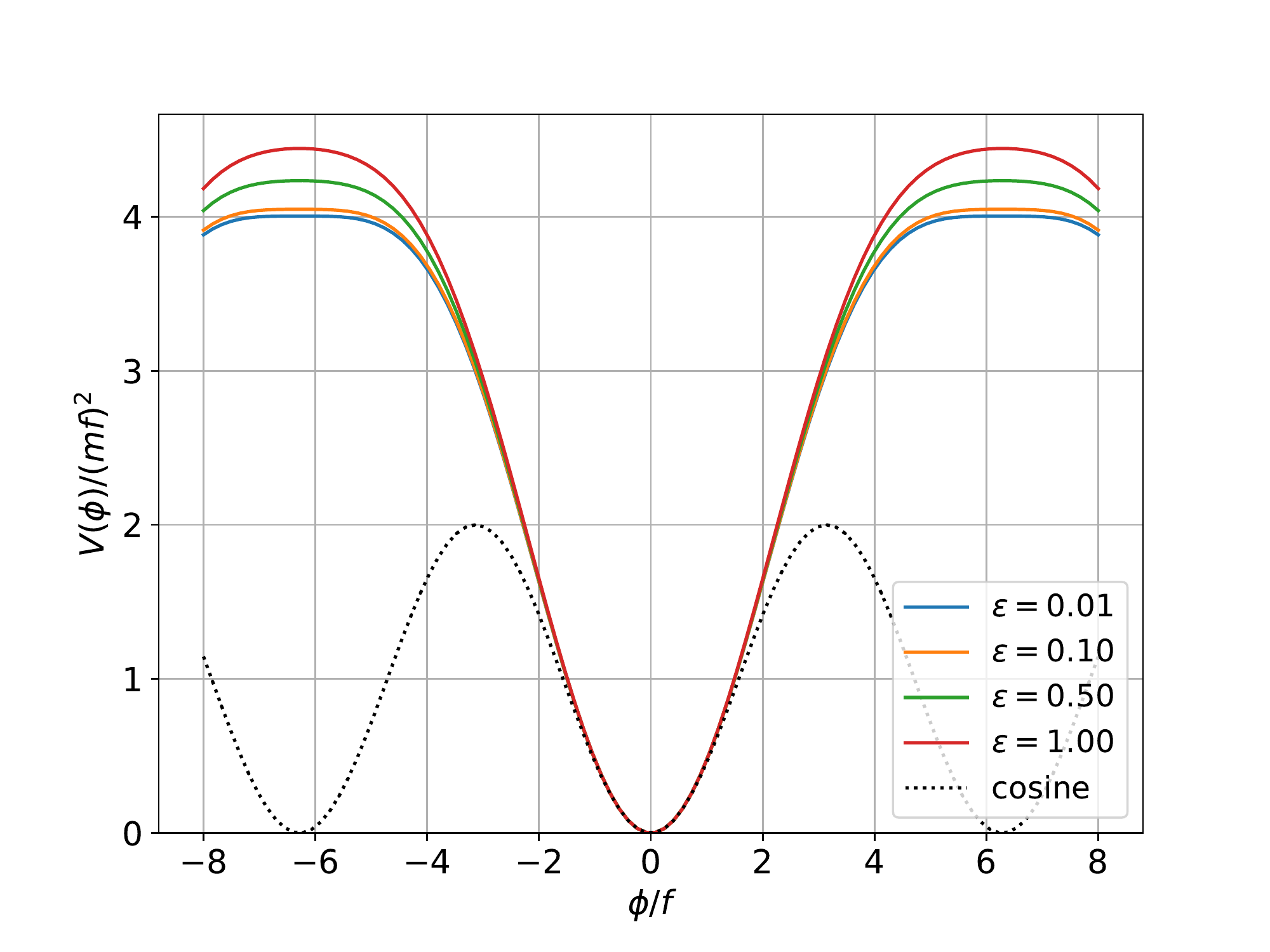}
    \caption{This plot shows the potential (\ref{eq:pot_align}) for $\epsilon=0.01,0.1,0.5,1.0$. The vertical axis denotes $V/(m(T)f)^2$. The black dotted line shows the cosine potential, given in Eq.~(\ref{eq:potcos}), which corresponds to $\epsilon= -4$. }
    \label{fig:cosalign}
\end{figure}

Here and in what follows, generalizing a single cosine potential, we consider two cosine terms whose amplitudes are given by $\Lambda_1^4 = \Lambda_2^4/(4+\epsilon) = m^2(T)f^2$ and the periods are given by $f_1 = f_2/2 = f$ as 
\begin{equation}
    V(\phi)=\frac{m^2(T) f^2}{2(1+\epsilon/8)} \left[5+\epsilon-\cos\left(\frac{\phi}{f}\right)-(4+\epsilon)\cos\left(\frac{\phi}{2f}\right)\right]
    % \tilde{V}(\tilde{\phi})=\left(\frac{m(T)}{m_{0}}\right)^2\frac{1}{2(1+\epsilon/8)}\left[5+\epsilon-\cos\tilde{\phi}-(4+\epsilon)\cos\frac{\tilde{\phi}}{2}\right].
    \label{eq:pot_align}
\end{equation}
with a constant parameter $\epsilon$ and $\Theta_1 = \Theta_2 = 0$. With this choice, the potential satisfies
\begin{align}
    V(0)= 0 \,, \qquad V (\phi) \sim \frac{m^2(T)}{2} \phi^2 \quad (|\phi| \to 0)\,,
\end{align}
where the second condition ensures that the ALP can behave as dark matter at a later time. When we set $\epsilon=-4$, the potential (\ref{eq:pot_align}) corresponds to the single cosine case given by Eq.~(\ref{eq:potcos}). Figure \ref{fig:cosalign} shows the potential shape of Eq.~(\ref{eq:pot_align}) for several values of $\epsilon$. It should be emphasized that, in the limit $\epsilon \to 0$, the quadratic term in the power series of cosine terms vanishes around the maximum of the potential and the quartic term becomes important. In that case, the potential hill becomes flatter as one can see in Fig.~\ref{fig:cosalign}.

Here we consider the temperature dependence of the ALP mass since interactions between the ALP and hidden sector gauge fields can in general introduce such a temperature dependence. Likewise the QCD axion, we parametrize the temperature dependence of the ALP mass as
\begin{equation}
    m^2(T)=\frac{m^2_0}{1+(T/T_c)^b},
\end{equation}
where $m_0$ denotes the ALP mass evaluated at $T=0$ and $T_c$ and $b$ are parameters depending on the hidden sector physics. In particular, $T_c$ can be different from the QCD confinement scale.

\subsection{Field evolution}
In the flat-FLRW Universe, the evolution equation of the axion is given by
\begin{equation}
    \frac{\partial^2}{\partial\tilde{t}^2}\tilde{\phi}+3\frac{H}{m_{\ast}} \frac{\partial}{\partial\tilde{t}} \tilde{\phi}-\frac{\partial^2_{\tilde{\sbm{x}}}}{a^2}\tilde{\phi}+\tilde{V}_{\tilde{\phi}}=0, \label{eq:KG}
\end{equation}
where $\tilde{t}\equiv m_{\ast}t$, $\tilde{\bm{x}}\equiv m_{\ast}\bm{x}$, $\tilde{V}=V/(m_\ast f)^2$, $\tilde{V}_{\tilde{\phi}}=d\tilde{V}/d\tilde{\phi}$  and $a$ is the scale factor. In the radiation dominated Universe with $a\propto t^{1/2}$, the Hubble parameter is given by $H/m_*=1/(2\tilde{t}) \propto T^2$. 

For the homogeneous mode of $\tilde{\phi}$, the evolution equation is given by
\begin{equation}
    \frac{d^2}{d\tilde{t}^2}\tilde{\phi}+3\frac{H}{m_{\ast}}\frac{d}{d\tilde{t}}\tilde{\phi}+\tilde{V}_{\tilde{\phi}}=0.\label{eq:KGhomo}
\end{equation}
When the axion mass is much smaller than the Hubble parameter, the axion field remains a constant value. On the other hand, when the temperature becomes lower than $T_\ast$ and the mass of the axion becomes larger than the Hubble parameter, the axion starts to oscillate. 
Perturbing Eq.(\ref{eq:KG}), the evolution equation of the linear perturbation $\delta\tilde{\phi}\equiv\delta\phi/f$ reads
\begin{equation}
    \frac{d^2}{d\tilde{t}^2}\delta\tilde{\phi}_k+3\frac{H}{m_{\ast}}\frac{d}{d\tilde{t}}\delta\tilde{\phi}_k+\left(\left(\frac{k}{am_{\ast}}\right)^2+\tilde{V}_{\tilde{\phi}\tilde{\phi}}\right)\delta\tilde{\phi}_k=0,
    \label{eq:KGinhomo}
\end{equation}
where $\delta\tilde{\phi}_k$ denotes the Fourier mode of $\delta\tilde{\phi}$, $\tilde{V}_{\tilde{\phi}\tilde{\phi}}\equiv d^2\tilde{V}/d\tilde{\phi}^2$, and we neglect the metric perturbation. From the next section, we analyze clump formation, considering the QCD axion and ALPs in turn.

%%%%%%%%%%%%%%%%%%%%%%%%%%%%%%
%%%%% Sec. SINGLE COSINE %%%%%
%%%%%%%%%%%%%%%%%%%%%%%%%%%%%%

\section{Clump formation: QCD axion}
\label{sec:clump_QCD}
In this section, we discuss the possibility of the clump formation for the QCD axion which commences the oscillation at the QCD epoch, taking into account the thermal correction to the axion mass~\cite{Borsanyi:2016ksw}.

\subsection{Linear calculation}  \label{SSec:QCDL}
When the misalignment axion is initially located around the top of the potential, it rolls down the region where the potential curvature, given by $\tilde{V}_{\tilde{\phi}\tilde{\phi}}=(m(T)/m_{\ast})^2\cos\tilde{\phi}$, becomes negative before the onset of the oscillation. Then, the the coefficient of $\delta\tilde{\phi}_k$ in Eq.(\ref{eq:KGinhomo}) can be negative for the modes that satisfy
\begin{equation}
    \frac{k}{am_{\ast}}< \frac{m(T)}{m_{\ast}}\label{eq:tachycond}.
\end{equation}
Using Eqs.(\ref{eq:KGhomo}) and (\ref{eq:KGinhomo}), the temperature dependence of the axion mass is given by 
\begin{equation}
    %\left(\frac{m(T)}{m_{\ast}}\right)^2=\frac{1}{(f/10^{12}{\rm GeV})^2( m_{\ast}/{\rm neV})^2}\frac{(7.6\times10^{-2})^4}{1+(T/T_c)^b}.
    \left(\frac{m(T)}{m_{\ast}}\right)^2=\left(\frac{f}{10^{12}{\rm GeV}}\right)^{-2}\left(\frac{ m_{\ast}}{10^{-9}{\rm eV}}\right)^{-2}\frac{(7.6\times10^{-2})^4}{1+(T/T_c)^b}.
\end{equation}
The Fourier mode of the fluctuation that satisfy Eq.~(\ref{eq:tachycond}) grows exponentially due to tachyonic instability. As discussed in Ref.~\cite{Fukunaga:2019unq}, the parametric resonance instability hardly takes place for the cosine potential. Furthermore, the time variation of $m$ makes the resonance even more difficult, since the periodicity of the oscillation changes in time. Therefore, the efficiency of the tachyonic instability is an important factor for the clump formation. As a situation where we expect the maximum enhancement through the tachyonic instability, we consider the case where the axion was initially located around the hilltop of the cosine potential, i.e., $\tilde{\phi}_i \simeq \pi$. An interesting example where tachyonic instability leads to formation of clumpy structure is Q-ball formation, studied e.g., in Refs.~\cite{Coleman:1985ki, Kusenko:1997si, Kasuya:1999wu} (see also Ref.~\cite{Hiramatsu:2010dx}).

\begin{figure}
\centering
\includegraphics[height=5.5cm%,width=.48\textwidth
]{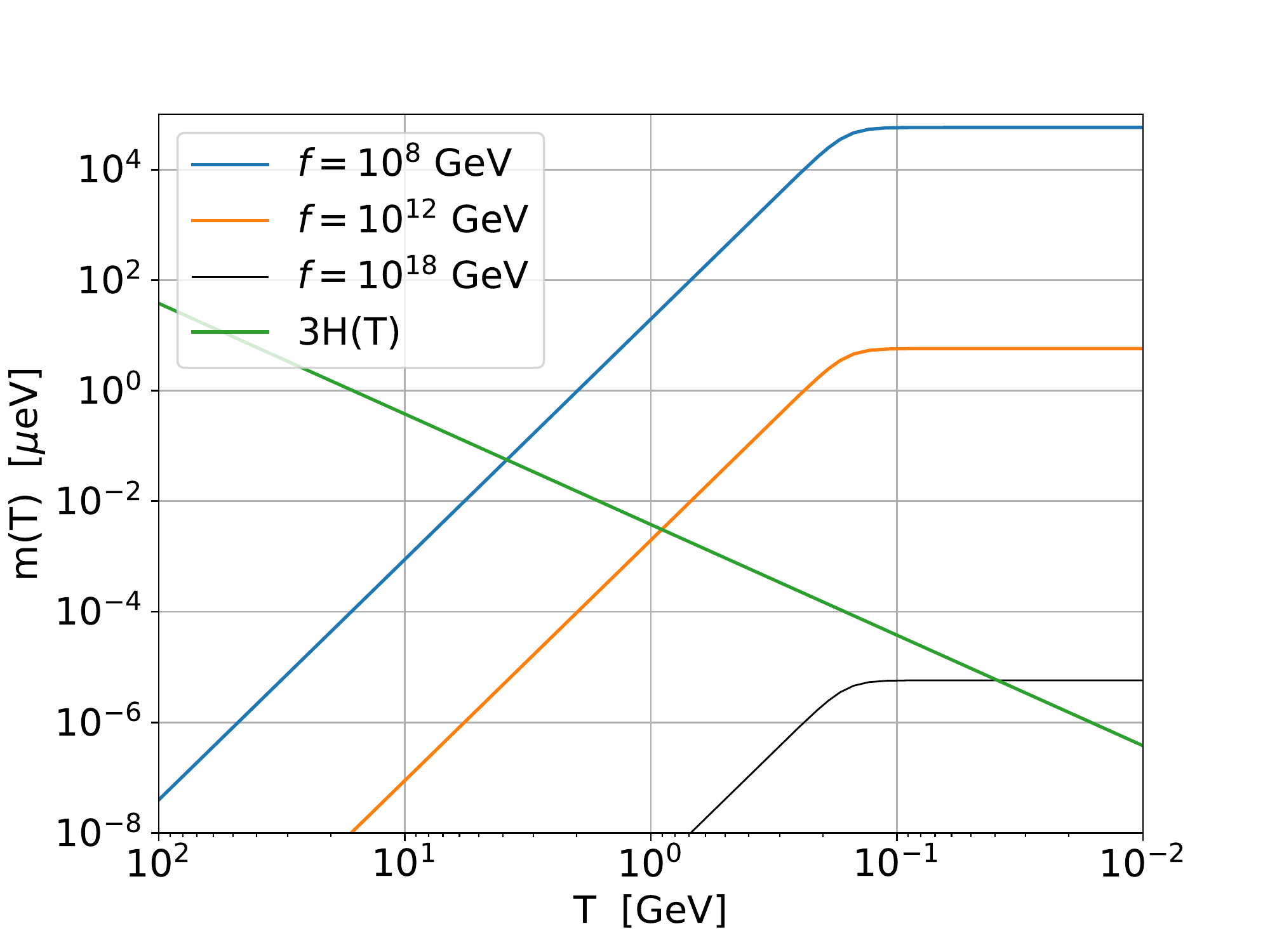}
\caption{\label{fig:decayconstdep2} This plot shows the $T$ dependence of the Hubble parameter $H$ and the axion mass $m$.}
\end{figure}
For a given decay constant $f$, the temperature $T_\ast$ with $3 H (T_\ast) = m (T_\ast)$ is uniquely determined. Figure~\ref{fig:decayconstdep2} shows the evolution of the Hubble parameter $H$ and the axion mass $m$ for $f=10^8$ GeV (blue) and 
$f=10^{12}$ GeV (orange). Notice that as long as the phenomenological constraint on $f$, given in Eq.~(\ref{Eq:rangef}), is satisfied, the QCD axion commences the oscillation in the QCD epoch, taking $T_\ast \sim {\cal O}({\rm GeV})$. For a comparison, we also show the axion mass for $f= 10^{18}$ GeV, for which, the axion mass has already reached $m_0$ at $T= T_\ast$.

The Fourier modes in the range (\ref{eq:tachycond}) exponentially grow as $\delta \tilde{\phi}_k \propto e^{\mu_k m_\ast t}$ with the dimensionless growth rate $\mu_k$, which is roughly estimated as
\begin{equation}
    \mu_k\simeq\sqrt{\left(\frac{m(T)}{m_{\ast}}\right)^2 - \left(\frac{k}{am_{\ast}}\right)^2}
\end{equation}
for $\tilde{\phi} \sim \pi$. For a smaller $k$, the growth rate becomes larger. In the limit $k/a \ll m(T)$, $\mu_k$ is given by $\mu_k \simeq m(T)/m_{\ast}$, implying that the growth rate $\mu_k$ becomes larger and larger as $m(T)$ increases.

The clump formation becomes more probable as $\tilde{\phi}_i$ is closer to $\pi$ and as the initial amplitude of the fluctuation $\delta \phi_i$ is larger. Meanwhile, when the amplitude of the initial fluctuation $\delta \phi_i$ is larger than the initial deviation from the top of the potential $|\tilde{\phi}_i - \pi|$, one spatial patch rolls down towards $\tilde{\phi} =0$ and another towards $\tilde{\phi} =2 \pi$, leading to formation of domain walls. Therefore, an initial condition where the clump formation is the most probable but the domain wall formation can be marginally avoided, we employ
\begin{align}
    \tilde{\phi}_i=\pi\times(1-10^{-8})\,, \qquad 
    \delta\tilde{\phi}_i=\pi\times10^{-8}\,. \label{IC}
\end{align}
The initial velocity of $\tilde{\phi}$ is determined by imposing the slow-roll condition as
\begin{align}
    \frac{d\tilde{\phi}}{d\tilde{t}} \bigg|_{\tilde{t}=\tilde{t}_i}=-\frac{\tilde{V}_{\tilde{\phi}}(\tilde{\phi}_i)}{3H_i/m_{\ast}}=-\left(\frac{T_{\ast}}{T_i}\right)^2\frac{7.6^4\times10^{-8}}{(f/10^{12}{\rm GeV})^2 (m_{\ast}/ 10^{-9}{\rm eV})^2(1+(T_i/T_c)^b)}\sin\tilde{\phi}_i.   \label{IC_v}
\end{align}

\begin{figure}
\centering
\includegraphics[height=5.5cm%,width=.48\textwidth
]{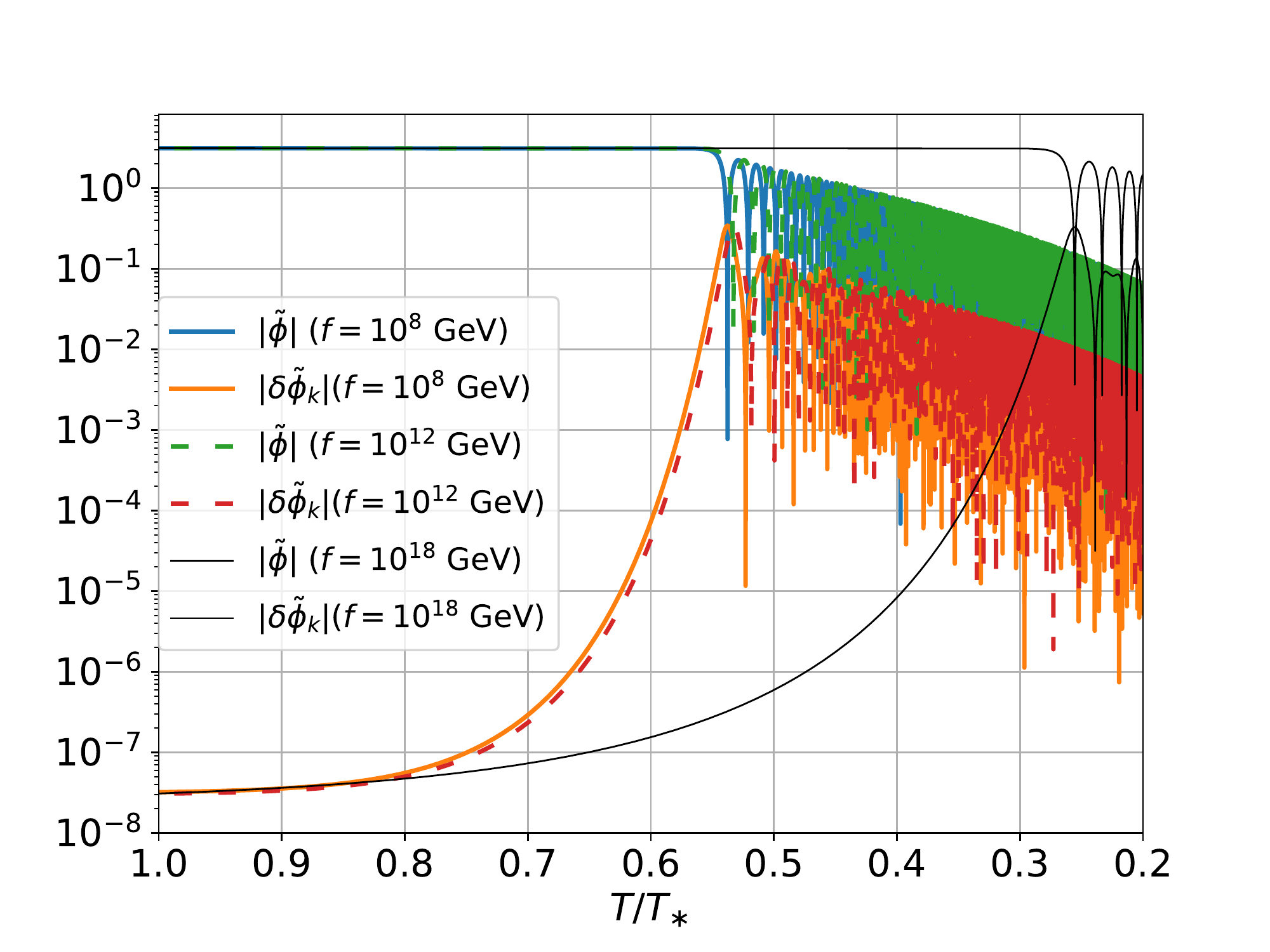}
\caption{\label{fig:decayconstdep} This panel shows the time evolution of the background homogeneous mode and the inhomogeneous mode with $k/(a_{\ast}m_{\ast})=0.1$ for $f=10^{8}$ GeV and $f=10^{12}$ GeV. The horizontal axis is the temperature normalized by $T_{\ast}$. For a comparison, we also show the time evolution for $f=10^{18}$ GeV, for which the axion mass remains time independent during the tachyonic instability.}
\end{figure}
Figure \ref{fig:decayconstdep} shows the time evolution of $\tilde{\phi}$ and $\delta\tilde{\phi}_k$. Here, the temperature of radiation is used as a corresponding time variable. The wavenumber $k$ is set to $k/(a_{\ast}m_{\ast})=0.1$, which satisfies the condition for the tachyonic instability, (\ref{eq:tachycond}). The temperature in the horizontal axis is normalized by $T_\ast$ to compare the evolution for different values of $f$, for which the axion commences the oscillation at different moments. For $f=10^{8}$ GeV and $f=10^{12}$ GeV, the period of the oscillation becomes shorter and shorter, since the axion mass still keeps on changing for a while after $T_\ast$. In both cases, the axion mass keeps on increasing as $m(T) \propto T^{-b/2}$ during the growth due to the tachyonic instability, the time evolution remains almost the same. The tachyonic instability terminates at the commencement of the oscillation.

For a comparison, in Fig.~\ref{fig:decayconstdep}, we have also plotted the evolution of $\tilde{\phi}$ and $\delta\tilde{\phi}_k$ for $f=10^{18}$ GeV, while it does not satisfy Eq.~(\ref{Eq:rangef}). In this case, as shown in Fig.~\ref{fig:decayconstdep2}, the axion starts to oscillate after $m(T)$ has reached the value at $T=0$. Since the smaller growth rate is compensated by the longer duration of the tachyonic instability, the total enhancement for $f=10^{18}$ GeV turns out to be almost the same as the one for $f=10^{8}$ GeV and $f=10^{12}$ GeV.

\subsection{Non-linear calculation}\label{sec:singlecos_nonlinear}
%Box size, gradient term
In the previous subsection, we computed the evolution of the inhomogeneous mode based on linear analysis. As a consequence of the exponential growth due to the tachyonic instability, the fluctuation ceases to be negligibly small when the initial condition is tuned around the top of the potential. In this subsection, we consider the nonlinear dynamics, solving Eq.~(\ref{eq:KG}) with the use of the lattice simulation. We set the number of grid points per edge, $N$, to $N=128$. Correspondingly, the cubic simulation box includes $128^3$ points on the lattice.

Because of the limited dynamic range in the lattice simulation, one has to carefully choose the box size and the grid number, which determine the minimum wavenumber $k_{\rm min}$ and the maximum wavenumber $k_{\rm max}$. In particular, the dynamic range should be properly determined in order to follow the tachyonic instability for the modes satisfying Eq.~(\ref{eq:tachycond}) and the subsequent subhorizon-scale dynamics such as the oscillon formation. Using the comoving box size of the lattice simulation, $L$, the maximum and minimum wavenumbers are given by $k_{\rm min} = 2 \pi/L$ and $k_{\rm max} = \pi N/L$.

% Hereafter, we use $N$ as the number of grid points per edge, i.e., the total number of points on the lattice is $N^{{\rm dims}}$ where dims means the number of dimensions, and $L$ is the size of the box. 

%The initial field configuration is given by the Gaussian distribution, characterized by mean $\mu$ and standard deviation $\sigma$. As a corresponding initial condition to Eq.~(\ref{IC}), we set
For computational simplicity, we assume that the initial field value for each grid point is determined by the Gaussian distribution, with the mean $\mu$ and the standard deviation $\sigma$. As a corresponding initial condition to Eq.~(\ref{IC}), we set
\begin{align}
    \mu=\pi\times(1-10^{-8})\,, \qquad \sigma=\pi\times10^{-8}\,. 
\end{align}
%When the initial field values for all the grid points on the lattice are determined simply based on the Gaussian distribution, the initial spectrum corresponds to a white noise for which the Fourier modes in the range
For a later use, let us introduce the minimum and maximum wavenumbres of the initial spectrum as $k_{\rm min, i}$ and $k_{\rm max, i}$, respectively. This initial condition corresponds to the white noise in the dynamic range, i.e., each Fourier mode has the same amplitude from $k_{\rm min, i}= k_{\rm min}$ till $k_{\rm max, i}= k_{\rm max}$. 
% $$
%   %\frac{2\pi}{L} \leq k \leq \frac{2\pi}{L} N,
%   \frac{2\pi}{L} \leq k \leq \frac{\pi}{L} N,
% $$
% i.e. the amplitude of each Fourier mode of $\delta\tilde{\phi}$ is scale-invariant.
%$$
% \sigma^2(\tilde{t}_i) = \sum_{p=1}^N \sum_{q=1}^N \sum_{r=1}^N |\delta \phi_{k_x= %2\pi p/L, k_y= 2\pi q/L, k_z= 2\pi r/L}(\tilde{t}_i)|^2 = |\delta %\phi(\tilde{t}_i)|^2 \times N^3\,.
%$$

%Since the tachyonic instability takes place only for the modes which satisfy Eq.~(\ref{eq:tachycond}), the resultant dynamics is sensitive to the choice of the minimum wavenumber $k_{\rm min}$ and the maximum wavenumber $k_{\rm max, i}$ of the initial spectrum. 
%We conduct the lattice simulation for different values of $k_{\rm min}$ by changing the box size $L$ and also for different values of $k_{\rm max, i}$ by assigning the same field values for neighbor $d^3$ grids, corresponding to $k_{\rm max, i} = 2 \pi/L \times N/d$. Notice that for $d \neq 1$, $k_{\rm max, i}$ does not coincide with $k_{\rm max}$, determined by the resolution limit, \YU{while $k_{\rm min}$ all the time coincides with the minimum wavenumber of the simulation, since the comoving simulation box is considered. {\bf Is this correct???} }

\begin{figure}
    \centering
    \includegraphics[height=5.5cm]{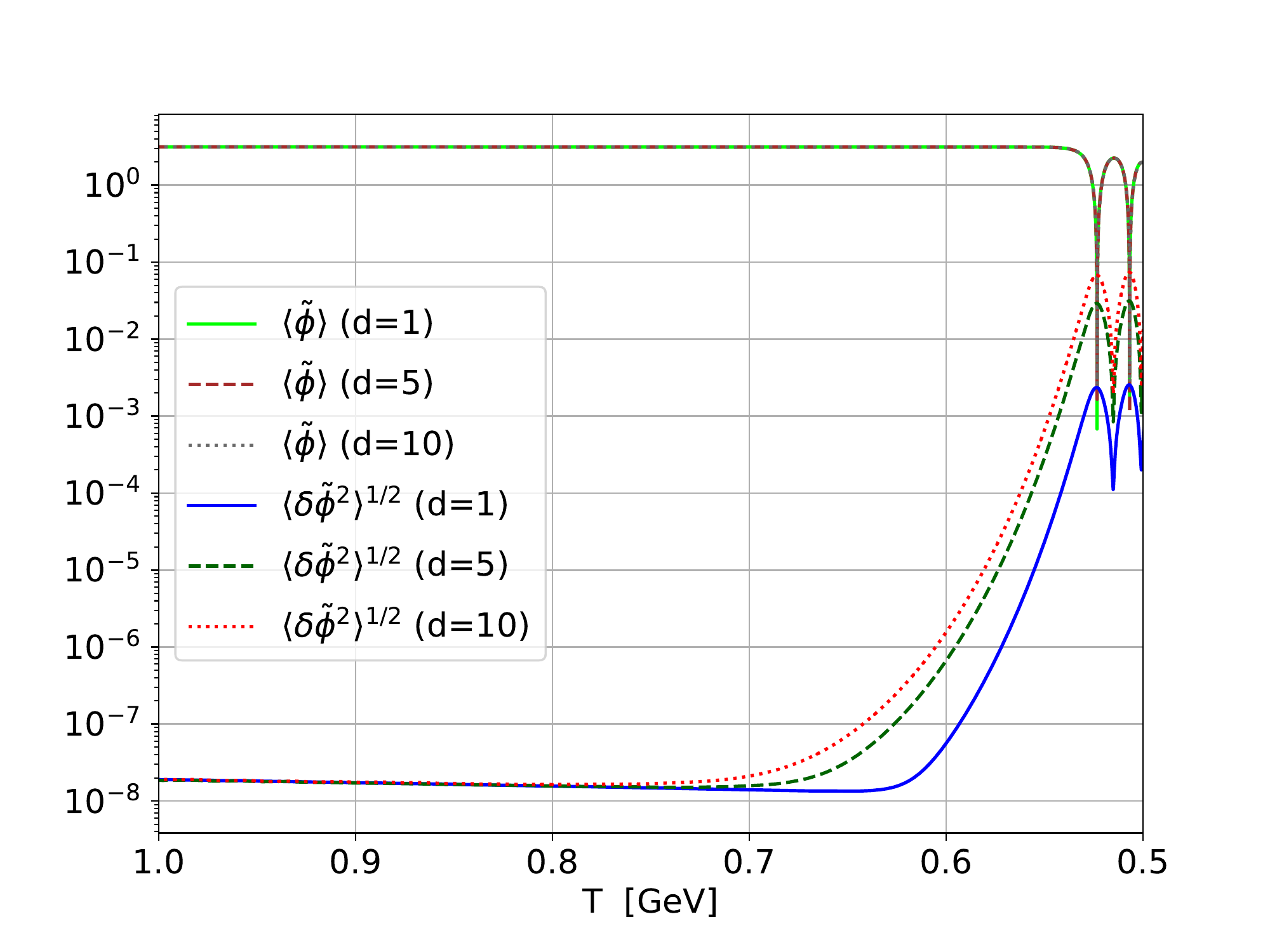}
    \caption{This plot shows the evolution of the mean and variance value. We chose the three different values of $d$ as $d=1$ (solid line), 5 (dashed line), and 10 (dotted line). Here, we chose $L=2\pi/(m_{\ast} a_{\ast})$. }
    \label{fig:NLsplitnum}
\end{figure}
Figure \ref{fig:NLsplitnum} shows the evolution of the homogeneous mode $\langle\tilde{\phi}\rangle$ and the root mean square of the fluctuation,  $\langle\delta\tilde{\phi}^2\rangle^{1/2}$, where the brackets express the spatial average. Here, the simulation box size is set to $L=2\pi/(m_{\ast} a_{\ast})$, corresponding to $k_{\rm min}= m_\ast a_\ast$. While keeping $k_{\rm min, i}= k_{\rm min}$, we chose several different values of $k_{\rm max, i}$ by assigning the same field values for neighbor $d^3$ grids with $d=1, 5, 10$. This corresponds to introducing a UV cut-off of the initial spectrum at $k_{\rm max, i} = \pi/L \times N/d$. For a given $\sigma$, the initial amplitude of the field fluctuation $\delta \phi$ is given by $|\delta \phi| \sim \sigma/({\rm number~of~the~modes})^{1/2}$ with the number of the modes being roughly $(k_{\rm max, i}/k_{\rm min, i})^3\sim (N/d)^3$. Therefore, the maximum amplitude of the fluctuation increases as we increase $d$, correspondingly as we decrease $k_{\rm max, i}$, while the homogeneous mode is independent of $d$. 
%\NK{This is because of our choice of the initial condition (Gaussian white noise with the fixed standard deviation) and can be removed if one impose the initial spectrum independent of the grid size. Thus, the result is not sensitive to the artificial high-k cutoff $k_{\rm max.i}$ and the validity of our simulation is justified.}
As shown in Fig.~\ref{fig:NLsplitnum}, once the axion has started the oscillation, the exponential growth due to the tachyonic instability terminates, because the amplitude of the axion rapidly decreases due to the Hubble friction (a more detailed analysis can be found in Ref.~\cite{Fukunaga:2019unq}). As a result, before the inhomogeneity reaches ${\cal O}(1)$, the axion is settled down around the potential minimum, where the deviation from the quadratic potential is negligible.

\begin{figure}
    \centering
    \includegraphics[height=5.5cm]{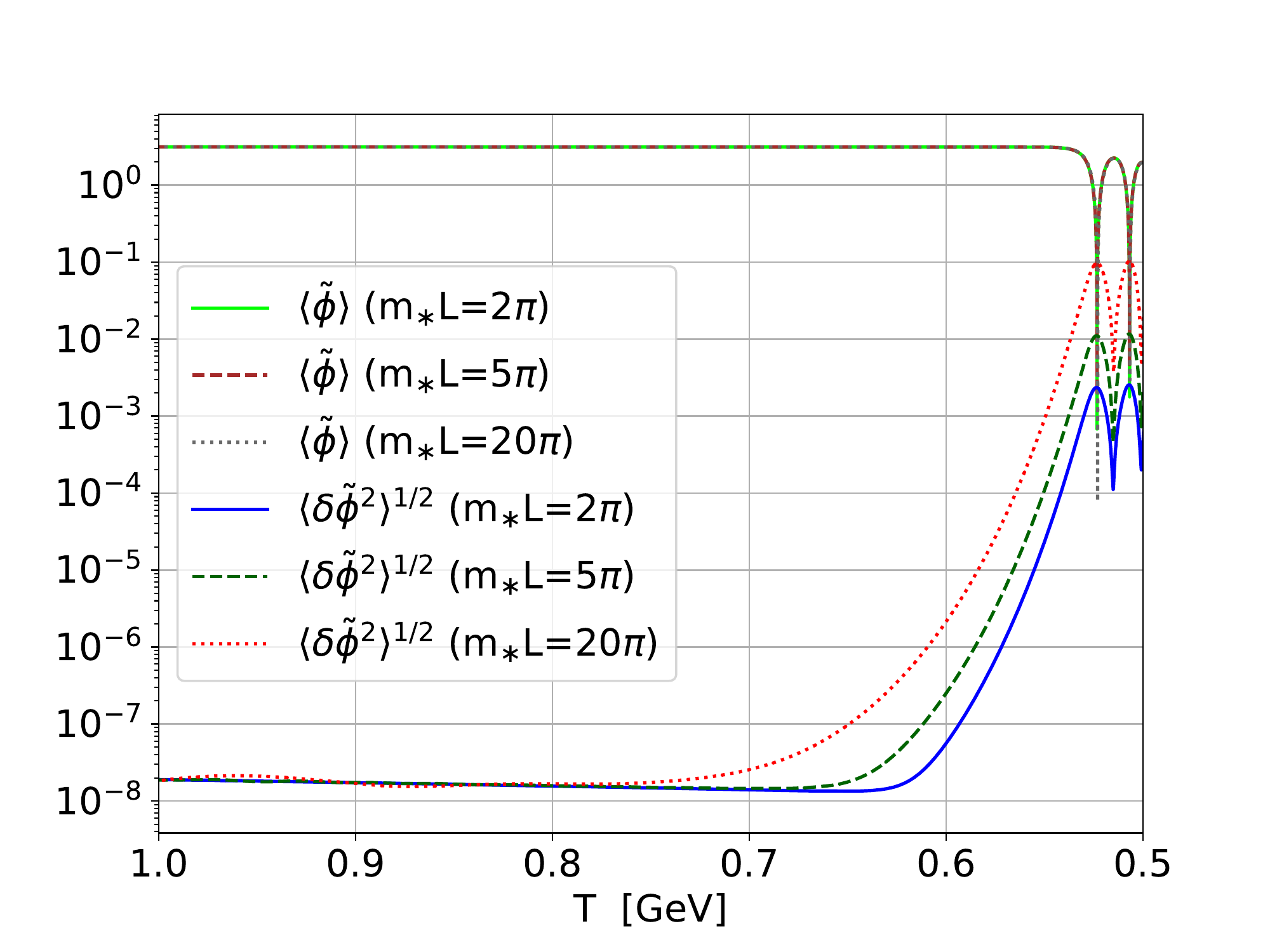}
    \caption{This plot shows the evolution of the mean and the variance value. We chose the three box size as $m_{\ast}L=2\pi$ (solid line), $m_{\ast}L=5\pi$ (dashed line), and $m_{\ast}L=20\pi$ (dotted line). }
    \label{fig:NLboxsize}
\end{figure}
In Fig.~\ref{fig:NLboxsize}, choosing $k_{\rm max, i} = k_{\rm max}= \pi N/L$, we have changed $k_{\rm min, i} = k_{\rm min}$ as $k_{\rm min, i}/(m_\ast a_\ast)=1,\, 2/5, 1/10$ by choosing $m_\ast a_\ast L = 2 \pi,\, 5 \pi,\,20 \pi$, respectively. As discussed in the previous subsection, the spatial gradient term disturbs the tachyonic instability, reducing the growth rate $\mu_k$. As we increase the box size, more and more low-$k$ modes, which undergo the tachyonic instability with the maximum growth rate $\mu_k \sim m(T)/m_\ast$ start to be included in the simulation. Therefore, the enhancement due to the tachyonic instability becomes more prominent for a larger box size. The $L$ dependence disappears, when we choose a sufficiently large $L$ so that the majority of the modes included in the simulation undergo the tachyonic instability, verifying that the result should not be altered by the change of the simulation setup.

Even for the largest simulation box with $m_\ast L = 20 \pi$, where the tachyonic instability becomes the most prominent, the instability has finished before the clump formation, being disturbed by the cosmic expansion. Here, we have chosen the size of the simulation box so that the minimum wavenumber becomes at most comparable to the Hubble scale. For example, for $m_\ast L = 20 \pi$, $k_{\rm min}$ amounts to $k_{\rm min}/(a_\ast H_\ast) = 0.3$. When we consider the super Hubble fluctuations, the metric perturbations, which are ignored in our computation, should be carefully considered.

\subsection{Uncertainty at high temperature}
In the previous subsection, we considered the cosine potential, which is predicted based on the DIGA. In Refs.~\cite{Borsanyi:2015cka, Borsanyi:2016ksw, Taniguchi:2016tjc, Petreczky:2016vrs}, it was shown that the DIGA well reproduces the lattice result in the high temperature range roughly above $T_c$. Meanwhile, it is known that for $T \geq {\cal O}(1)$ GeV, the lattice computation becomes rather challenging, because of the difficulty in sampling topologically non-trivial configurations. See Refs.~\cite{Borsanyi:2016ksw, Frison:2016vuc, Taniguchi:2016tjc, Jahn:2018dtn} for the recent studies about the lattice simulation in the high temperature range. Because of that, for $T \geq {\cal O}(1)$ GeV, the temperature dependence and the potential form of the axion have not been clearly understood as much as for $T < {\cal O}(1)$ GeV. Having considered this, in this subsection, we investigate whether the axion clump can be formed or not, when the temperature dependence of the axion mass or the potential form is modified for $T \geq {\cal O}(1)$ GeV.

First, we consider the case where the temperature dependence of $m$ becomes different for $T \geq T_{\ast\ast} = 1.5$GeV as $m(T)\propto T^{-b_{\rm H}}$, while keeping $m(T)\propto T^{-b_{\rm L}}$ with $b_{\rm L} =8.7$ for $T \leq T_{\ast\ast} = 1.5$GeV. 
\begin{figure}
\centering
\includegraphics[height=5.5cm,width=.48\textwidth]{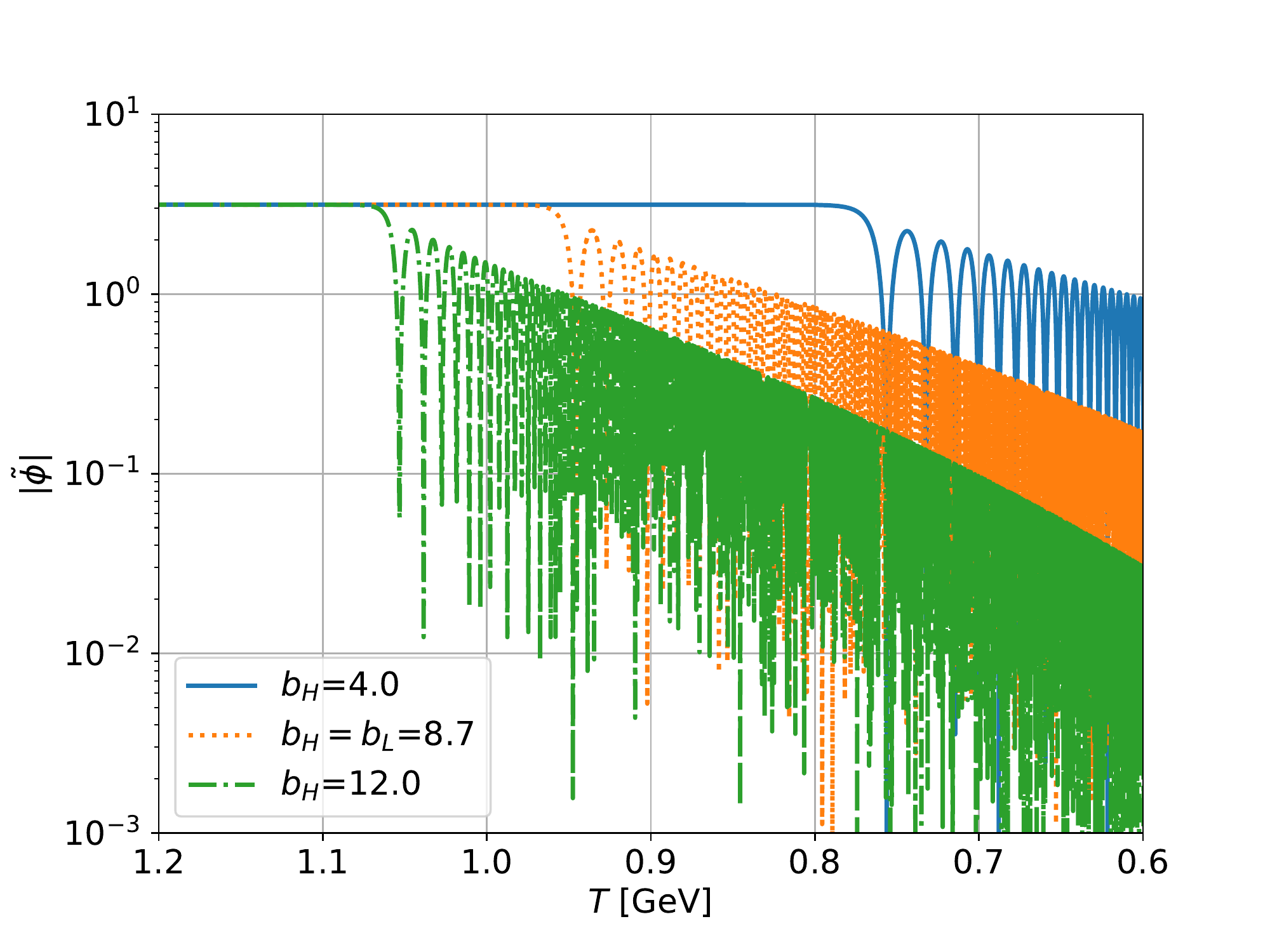}
\hfill
\includegraphics[height=5.5cm,width=.48\textwidth]{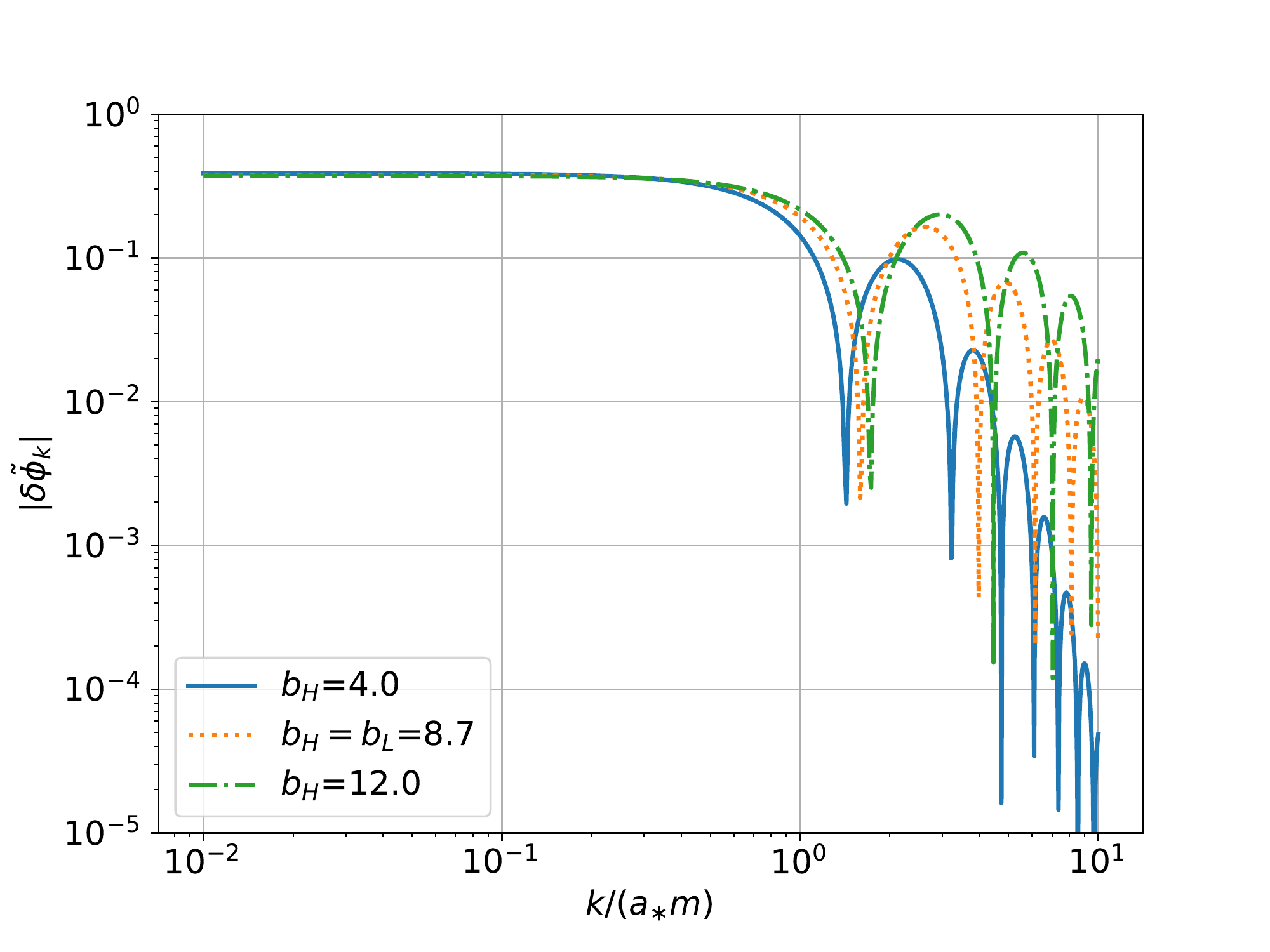}
\caption{\label{fig:Powtemp} The left panel shows the evolution of background homogeneous mode for different $b_H$. Here, we set $b_H=4.0,8.7,12.0$, $T_{\ast\ast}=1.5$GeV, and $f=10^{11}$GeV. The right panel shows the spectrum for different $b_H$ when $|\delta\tilde{\phi}_k|$ has reached the maximum amplitude.}
\end{figure}
Figure \ref{fig:Powtemp} shows the evolution of the background field $\tilde{\phi}$ and the amplitude of the linear perturbation $\delta \tilde{\phi}_k$ for $b_{\rm H} = 4.0,8.7,12.0$. Here we set the initial condition as Eqs.~(\ref{IC}) and (\ref{IC_v}) and the decay constant as $f=10^{11}$GeV. The amplitude of $\delta \tilde{\phi}_k$ is evaluated at $t= t_{\ast}$, where $|\delta \tilde{\phi}_k|$ has reached the maximum value through the tacyonic growth. In Sec.~\ref{SSec:QCDL}, we argued that the temperature dependence of $m(T)$ does not significantly change the net enhancement due to the tachyonic instability. As is expected from this, the maximum amplitude of $\delta \tilde{\phi}_k$ remains almost the same, even if we change the value of $b_{\rm H}$.

Our next target is considering the case where the potential form is modified from Eq.~(\ref{eq:potcos}) at $T \geq T_{\ast\ast} = 1.5$GeV. Here, instead of considering a modified potential directly, we consider a modification of the initial velocity as $d\tilde{\phi}/d\tilde{t}|_{\tilde{t}_i= \tilde{t}_{\ast \ast}}=d\tilde{\phi}/d\tilde{t}|_{{\rm SR}}\times 10^c$ choosing the initial time as $\tilde{t}_{\ast\ast}= \tilde{t}(T_{\ast \ast})$, where $d\tilde{\phi}/d\tilde{t}|_{{\rm SR}}$ denotes the initial velocity determined by imposing the slow-roll condition, (\ref{IC_v}). Choosing $c< 0$ ($c>0$) corresponds to considering the case where the axion was initially located at a shallower (steeper) potential region than Eq.~(\ref{eq:potcos}) under the slow-roll condition. 
\begin{figure}
\centering
\includegraphics[height=5.5cm,width=.48\textwidth]{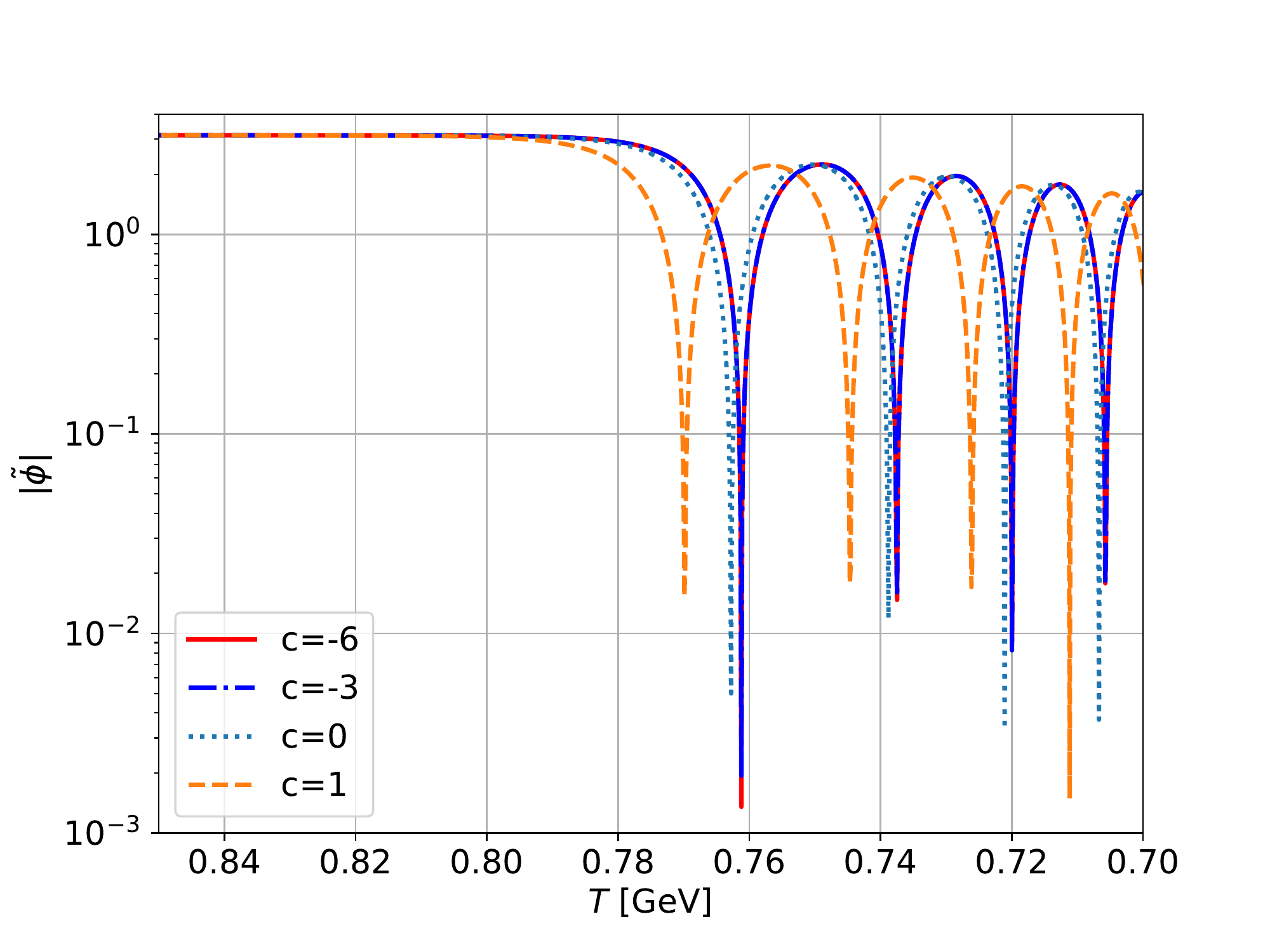}
\hfill
\includegraphics[height=5.5cm,width=.48\textwidth]{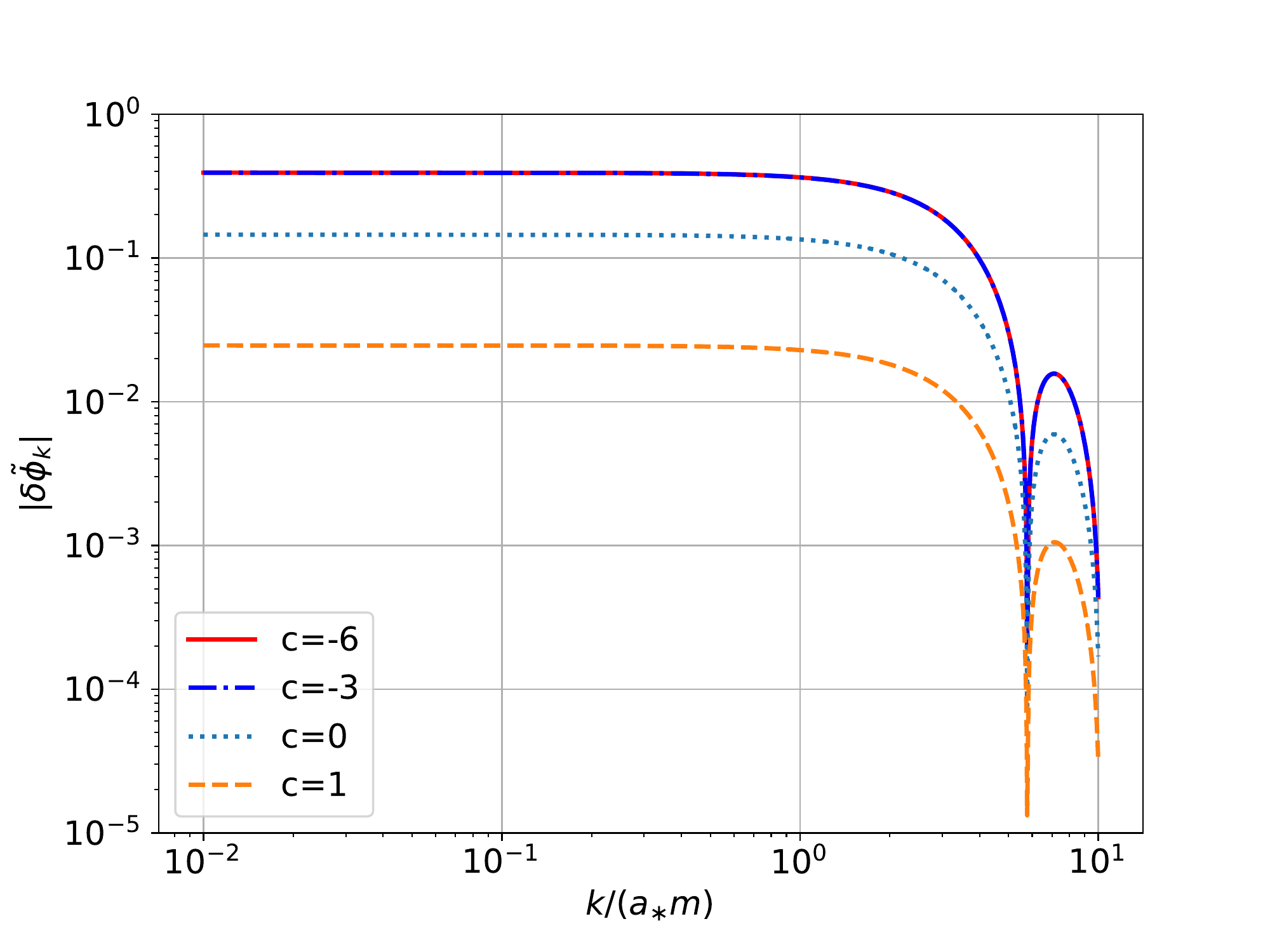}
\caption{\label{fig:SRbroken}The left panel shows the evolution of background homogeneous mode and the right panel shows the spectrum when $|\delta\tilde{\phi}_k|$ has reached the maximum amplitude. Here, we set $c$ to $c=-6,-3,0,1$. }
\end{figure}
Figure \ref{fig:SRbroken} shows the evolution of the background field $\tilde{\phi}$ and the amplitude of the linear perturbation $\delta \tilde{\phi}_k$ for $c = -6, -3, 0, 1$. Again, we set the initial condition as Eqs.~(\ref{IC}) and (\ref{IC_v}) and the decay constant as $f=10^{11}$GeV. For $c< 0$, the enhancement due to the tachyonic instability becomes slightly larger than $c=0$, addressed in the previous subsection. However, even with this modification, the tachyonic instability terminates before the inhomogeneity $|\delta \tilde{\phi}_k|$, which roughly amounts to $|\delta \phi/\phi|$, becomes ${\cal O}(1)$. This is because the sustainability of the tachyonic instability is mainly determined by the dynamics around $T = T_\ast$.

In this section, we have shown that the inhomogeneity of the axion grows exponentially due to the tachyonic instability. Nevertheless, the enhancement has turned out to be insufficient for the axion clump formation in the pre-inflationary scenario, verifying the prevailing understanding.

%%%%%%%%%%%%%%%%%%%%%%%%%%%%%%%%%%%%%
%%%%%% Sec. MULTIPLE COSINE %%%%%%%%%
%%%%%%%%%%%%%%%%%%%%%%%%%%%%%%%%%%%%%
\section{Clump formation: ALPs}
\label{sec:clump_ALPs}
In the previous section, focusing on QCD axion, we showed that the tachyonic instability is not enough for the clump formation in the scenario where the PQ symmetry breaking takes place before or during inflation. In this section, we investigate the possibility of the clump formation for ALPs, which can have a more general form of the potential. There are only few studies about ALP clump formation in the pre-inflationary scenario. In Ref.~\cite{Hardy:2016mns}, Hardy considered a scenario where a hidden sector from which an ALP acquires the mass undergoes the first order phase transition. As a consequence, it was argued that the ALP mass inside formed bubbles becomes larger than the one outside bubbles. The difference in the ALP mass results in the difference in the commencement time of the oscillation, generating the inhomogeneity of the energy density due to the difference in the dilution factor. In Refs.~\cite{Fukunaga:2019unq, Kawasaki:2019czd}, it was shown that when the ALP potential has an extensive shallow region likewise the potential proposed in Ref.~\cite{Nomura:2017ehb}, the parametric resonance becomes rather efficient, leading to ${\cal O}(1)$ inhomogeneity \cite{Fukunaga:2019unq} and subsequently to the formation of overdense clumps \cite{Kawasaki:2019czd}, called oscillons, also in the pre-inflationary scenario. Since the potential addressed in Refs.~\cite{Fukunaga:2019unq, Kawasaki:2019czd} is very different from the cosine potential, one may wonder whether a similar situation can be realized just with a milder modification from the conventional cosine potential. In this section, we investigate the possibility of ALP clump formation for a potential with multiple cosine terms, presented in \ref{sec:alignedpotential}.

\subsection{Linear calculation}
In this section, we investigate whether the ALP fluctuation can be saturated to the background homogeneous mode, considering the potential (\ref{eq:pot_align}) with several different values of $\epsilon$. When the value of $\epsilon$ is close to zero, the top of the potential becomes flatter (see $\epsilon=0.01$ in Fig.~\ref{fig:cosalign}.). Then, the low-$k$ modes of $\delta\tilde{\phi}_k$ can stay long in the tachyonic region, growing exponentially. Furthermore, when the ALP is located at a shallow potential region before the commencement of the oscillation, the parametric resonance sustains longer, since the delay of the onset of the oscillation makes the effect of the cosmic expansion less effective. Changing the initial condition and the potential parameter $\epsilon$, we consider the instabilities classified into the following three categories:
\[
|\delta{\phi}/\phi|\ {\rm becomes}\  {\cal O}(1)\  {\rm by} = 
 \begin{cases}
 {\rm tachyonic \ instability} &{\rm (i)} \\
 {\rm tachyonic\ instability\ \&\ parametric\ resonance} & {\rm (ii)} \\ 
 {\rm parametric \ resonance} & {\rm (iii)}
  \end{cases}
\]
one by one.

\begin{figure}
    \centering
    \includegraphics[height=5.5cm]{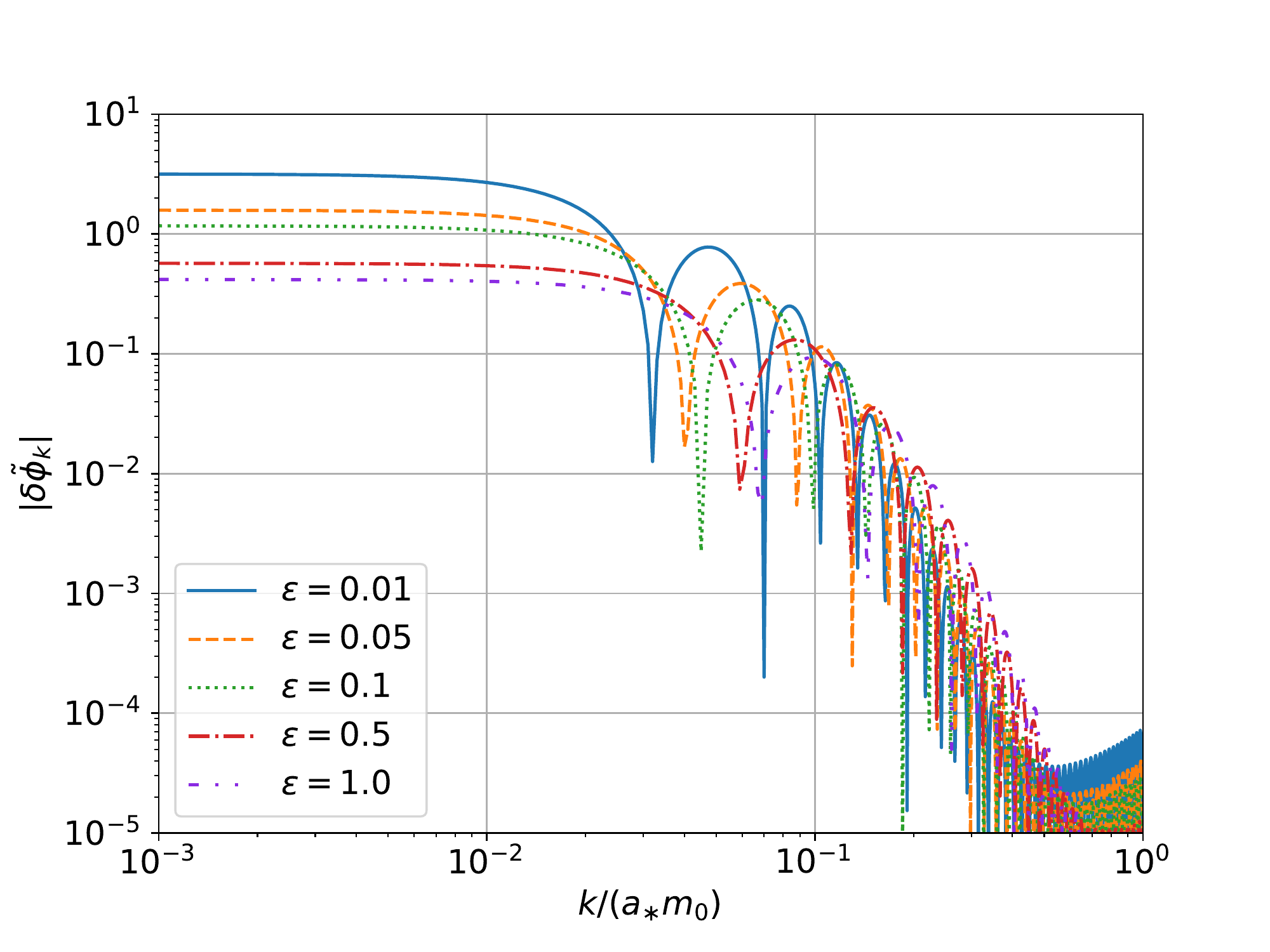}
    \caption{This plot shows the spectrum of the ALP field fluctuation for the potential (\ref{eq:pot_align}) with $\epsilon$=0.01 (blue solid), 0.05 (orange dashed), 0.1 (green dotted), 0.5 (red dash-dotted), and 1.0 (purple dash-dot-dotted), respectively. Each spectrum is evaluated, when the amplitude has reached the maximum value. Here, the initial condition is set as $\tilde{\phi}_i=2\pi\times(1-10^{-7})$ and $\delta\tilde{\phi}_{k, i}=2\pi\times10^{-8}$ at $\tilde{T}_i=5.0$. In this case, the ALP fluctuation is solely enhanced by the tachyonic instability (type (i)).
    For $\epsilon=0.01,\, 0.05,\, 0.1$, since $|\delta \tilde{\phi}_k|$, which roughly corresponds to $|\delta \phi/\phi|$, reaches ${\cal O}(1)$, the computation based on the linear analysis is not reliable. }
    \label{fig:speclinear_tac}
\end{figure}

\subsubsection{Type i: Tachyonic instability}
When we consider the case where $\phi$ %was located at an extensively shallow potential region 
is tuned around the top of the potential ($\tilde\phi \simeq 2\pi$) before the onset of the oscillation, the tachyonic instability can enhance the inhomogeneity of the ALP field. %to be ${\cal O}(1)$.
We have numerically followed the evolution of each Fourier mode of field fluctuation together with the background field. 
We start our calculation at $\tilde{T}_*=T_*/T_c$.\footnote{The following result remains almost the same, even if we start the computation in an earlier time.} 
In what follows, we express $\tilde{\phi}$ and $\delta \tilde{\phi}$ evaluated at $T= T_*$ as
\begin{align}
    \tilde{\phi}_i \equiv \tilde{\phi} (T_*)\,, \qquad \quad \delta \tilde{\phi}_{k, i} \equiv \delta \tilde{\phi}_k (T_*)
\end{align}
For $\tilde{T}_*\ll 1$ or equivalently $T_* \ll T_c$, the ALP mass remains constant. On the other hand, for $\tilde{T}_* \geq 1$ or $T_* \geq T_c$, $m(T)$ varies as the temperature decreases during the oscillation.

Figure~\ref{fig:speclinear_tac} shows the resultant spectrum of $\delta\tilde{\phi}_k$  for five different values of $\epsilon$, evaluated when the fluctuation has reached the maximum amplitude. Here, we set $b=8.0$ and employ the initial condition for the background and fluctuation as
\begin{align}
\tilde{\phi}_i=2\pi\times(1-10^{-7})\,, \qquad \delta\tilde{\phi}_{k,i}=2\pi\times10^{-8}\,, % \qquad \tilde{T}_* = 5
\end{align}
at $\tilde{T}_*=5.0$. Since the dominant instability is the tachyonic instability, all the low $k$ modes are uniformly enhanced. In the previous section, for the QCD axion with the cosine potential, we showed that the tachyonic instability terminates before $|\delta \phi/\phi|$ reaches ${\cal O}(1)$, since the axion field rapidly rolls down the potential, exiting the negative curvature region of the potential. For the potential (\ref{eq:pot_align}) with $\epsilon < 1$, since the gradient of the potential around the hilltop is shallower than the one for the cosine potential, the tachyonic instability continues longer. As a result, when $\tilde{\phi}_i$ is tuned around $\pi$, $|\delta \phi/\phi|$ can reach ${\cal O}(1)$ only through the exponential growth due to the tachyonic instability. For $\epsilon=0.01,\, 0.05,\, 0.1$, since $|\delta \tilde{\phi}_k|$, which roughly corresponds to $|\delta \phi/\phi|$, is enhanced to be ${\cal O}(1)$, we need to take into account the non-linearity to compute the dynamics properly.

When the ALP fluctuation is significantly enhanced by the tachyonic instability, we also need to make sure that this does not contradict to the isocurvature constraint in the CMB scales~\cite{Akrami:2018odb}. To solve the super Hubble evolution, we also need to take into account the metric perturbations, which are ignored in this paper. The impact of the self-interaction on the isocurvature constraint was discussed in Ref.~\cite{Kobayashi:2013nva}.

%As we discussed in Sec.~\ref{sec:singlecos_nonlinear}, however, the gradient term suppress the growth of the fluctuation. Then, it seems that the range of $\epsilon\ (0<\epsilon\leq 0.1)$ where apparent saturation occurs is narrowed by nonlinear calculation. 

\begin{figure}
    \centering
    \includegraphics[height=5.5cm,width=.48\textwidth]{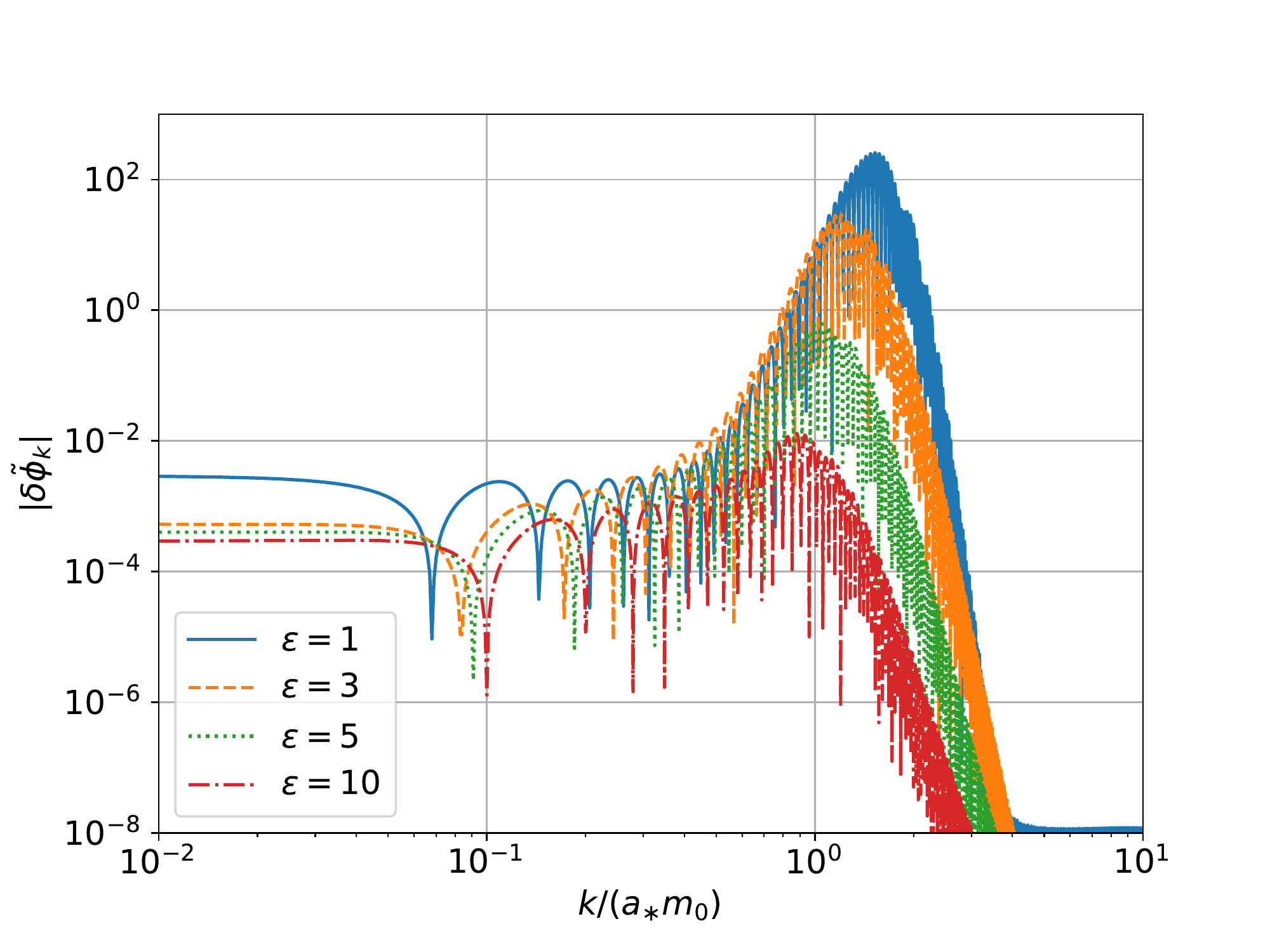}
    \hfill
    \includegraphics[height=5.5cm,width=.48\textwidth]{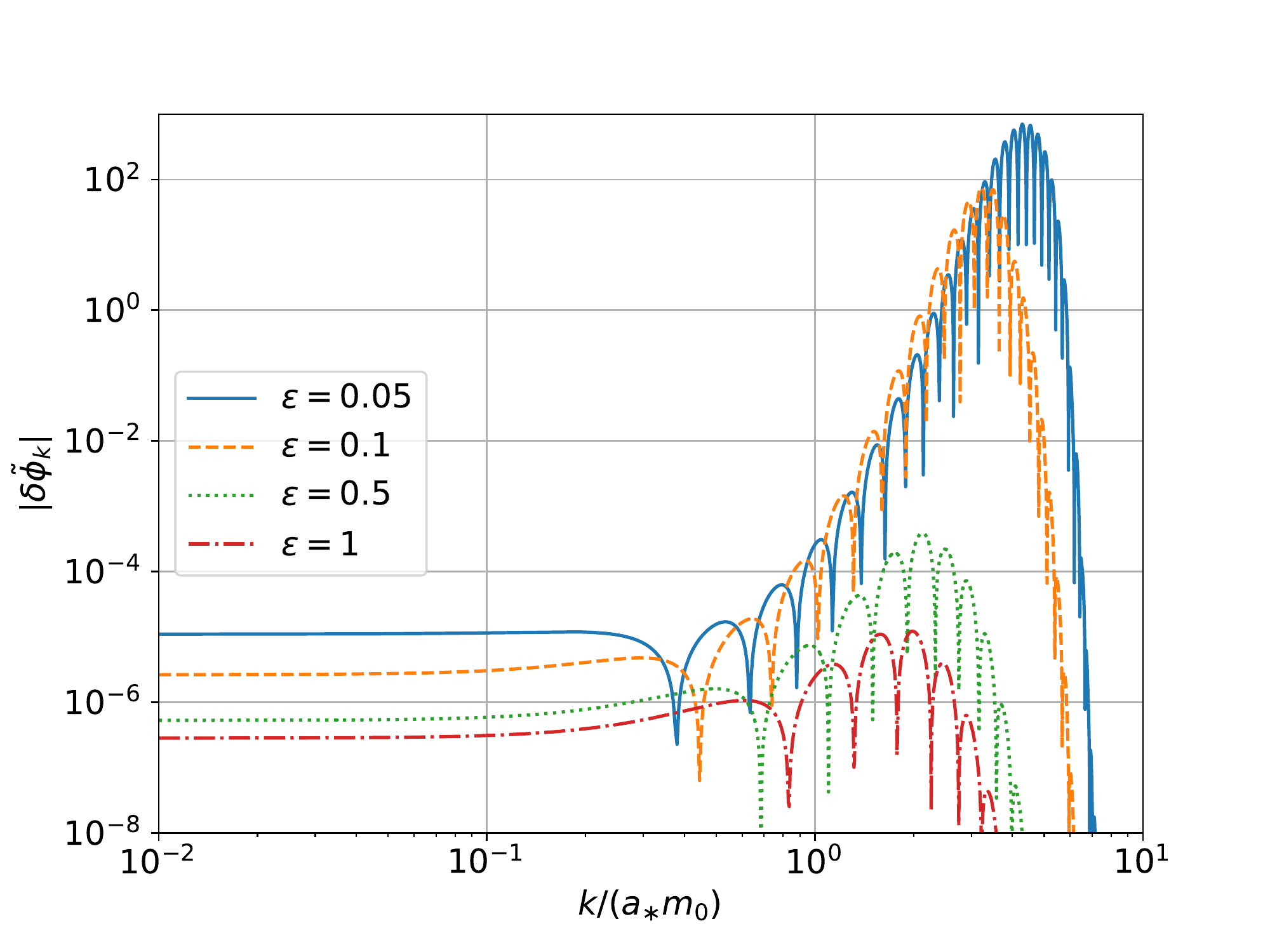}
    \caption{\label{fig:speclinear_param}The left panel shows the spectrum for $\epsilon$=1 (blue solid), 3 (orange dashed), 5 (green dotted), 10 (red dash-dotted). The initial condition is given by $\tilde{\phi}_i=2\pi\times(1-10^{-5})$ and $ \delta\tilde{\phi}_i=2\pi\times10^{-8}$ at $\tilde{T}_*= T_*/T_c=5.0$. In this case, the inhomogeneity becomes ${\cal O}(1)$ through the tachyonic instability and the succeeding resonance instability (type (ii)). The right panel shows the spectrum for $\epsilon$=0.05 (blue solid), 0.1 (orange dashed), 0.5 (green dotted), 1 (red dash-dotted). The initial condition is set by $\tilde{\phi}_i=2\pi\times(1-10^{-2})$ and $\delta\tilde{\phi}_{k, i}=2\pi\times10^{-8}$ at $\tilde{T}_i=0.5$. In this case, the inhomogeneity becomes ${\cal O}(1)$ solely through the resonance instability (type (iii)). For $\epsilon=5, 10$ in the left panel and for $\epsilon=0.5, 1$ in the right panel, the spectra are evaluated, when they have reached the maximum values, while for the rest, they are evaluated right after $|\delta \tilde{\phi}_k|$ has reached ${\cal O}(1)$, where the non-linear analysis is required. }
    %\caption{\label{fig:speclinear_param}The left panel shows the spectrum for $\epsilon=0.1$ and the right panel shows the one for $\epsilon=1.0$. The spectrum at different oscillation periods are shown by different colors. The wave number shown in the horizontal axis is divided by the scale factor at $3H=m$. }
\end{figure}

\subsubsection{Type ii: Tachyonic \& Parametric resonance instabilities}
%When $\phi$ was initially located as a region where the potential is not shallow enough, 
When $\phi_i$ is not fine-tuned around a shallow potential region as much as type (i), 
%or the potential hill is not sufficiently flat,  
the tachyonic instability finishes before the inhomogeneity becomes ${\cal O}(1)$. Nevertheless, in some cases, the tachyonic instability is followed by the resonance instability. In the end, the amplitude of $\delta \phi/\phi$ can reach ${\cal O}(1)$ through the tachyonic instability and the succeeding resonance instability.

The left panel of Fig.~\ref{fig:speclinear_param} shows the spectrum for $\epsilon=1$ (blue solid), $\epsilon=3$ (orange dashed), $\epsilon=5$ (green dotted), $\epsilon=10$ (red dash-dotted). The initial values of the background and each Fourier mode are set as 
\begin{align}
    \tilde{\phi}_i=2\pi\times(1-10^{-5})\,, \qquad \delta\tilde{\phi}_{k, i}=2\pi\times10^{-8}\,, 
\end{align}
at $\tilde{T}_* = 5$. After the low-$k$ modes are enhanced by the tachyonic instability, the resonance instability creates a peak in the spectrum corresponding to  the first resonance band. We will discuss the time evolution in more detail in the next subsection.  

% Since we set initial amplitude of $\tilde{\phi}$ as the deviation from the potential top becomes very small and the mass depends on the temperature, low-k mode is amplified due to tachyonic instability. Compared to the conventional cosine potential, the onset of the oscillation is delayed and damping due to cosmic expansion is reduced, parametric resonance occurs and spectrum has a peak. 

%Next, we investigate the type (ii). Figure~\ref{fig:speclinear_param} shows the spectrum for $\epsilon=0.5$(left panel), and $\epsilon=1.0$(right panel). In both cases, low-k modes are uniformly amplified by tachyonic instability. Compared to the conventional cosine potential, the onset of the oscillation is delayed, and damping due to cosmic expansion is reduced, then background homogeneous mode can climb to a region where the curvature of potential is negative. Then, tachyonic instability occurs more than one time. On the other hand, the spectrum has a peak at a certain wave number due to the parametric resonance. Since the characteristic amplitude of fluctuation of the length scale $k^{-1}$ is the value obtained by multiplying this spectrum by $k^3/(2\pi^2)$. Then, it is expected that the fluctuation can saturate to the homogeneous background mode combining tachyonic instability and parametric resonance when the value of $\epsilon$ is between 0.1 to 1.0.

\subsubsection{Type iii: Resonance instability}
In general, when the frequency of the oscillation, $\omega$, significantly changes in time, taking $|\dot{\omega}/\omega^2| \geq {\cal O}(1)$, a sustainable parametric resonance hardly takes place. Meanwhile, for $T_* \ll T_c$, $m(T)$ has already stopped evolving, when the ALP commences the oscillation. In that case, the parametric resonance can efficiently enhance the inhomogeneous mode of $\phi$ during the oscillation. In addition, when the oscillation starts much later than the time $t= t_\ast$ with $3H_\ast = m_\ast$, the time scale of the oscillation is much shorter than the one of the cosmic expansion already just after the commencement of the oscillation. In this case, the sustained parametric resonance can take place without being disturbed by the cosmic expansion. As discussed in Refs.~\cite{Kitajima:2018zco, Fukunaga:2019unq}, the delayed onset is characterized by 
\begin{align}
    H_{\rm osc} \sim \sqrt{|(dV(\phi_i)/d\phi_i)/\phi_i|} \quad {\rm or} \quad H_{\rm osc}/m \sim \sqrt{|(d \tilde{V}(\tilde{\phi}_i)/d\tilde{\phi}_i)/\tilde{\phi}_i|} \,,
\end{align}
when $\phi$ is not the dominant component of the Universe (see Ref.~\cite{Patel:2019isj} about the evaluation of $H_{\rm osc}/m$ when $\phi$ is the dominant component). Therefore, when $\phi$ was initially located at a potential region where the gradient is shallower than the one for the quadratic one, the oscillation starts much later than $t= t_\ast$, leading to the efficient parametric resonance.

%Parametric resonance becomes efficient when the mass of the axion is constant since the modes in resonance bands are less likely to shift than when the mass depends on temperature. 

The right panel of Fig.~\ref{fig:speclinear_param} shows the spectrum for $\epsilon=0.05$ (blue solid), 0.1 (orange dashed), 0.5 (green dotted), 1 (red dash-dotted). The initial condition is set as 
\begin{align}
    \tilde{\phi}_i=2\pi\times(1-10^{-2})\,, \qquad \delta\tilde{\phi}_{k, i}=2\pi\times10^{-8}\,, 
\end{align}
at $\tilde{T}_* = 0.5$. In these cases, while the tachyonic instability does not persist, the succeeding resonance instability leads to the ${\cal O}(1)$ inhomogeneity. As a result, the resonance instability creates a peak around the first resonance band. This case was addressed in Refs.~\cite{Fukunaga:2019unq, Kawasaki:2019czd}.  

%Finally, we investigate the possibility (iii). From the simple point of view, this can be realized by taking $\epsilon$ larger than (ii). However, when $\epsilon$ is increased, the delay of the onset of the oscillation decreases and the evolution of the homogeneous mode settles on harmonic oscillation earlier. Thus, in addition to less effective tachyonic instability, parametric resonance also becomes inefficient. 
%In the parameter region we investigated, there was no saturation that caused by the parametric resonance as the main factor to realize the situation (iii).
%On the other hand, parametric resonance is effective when the mass of the axion is constant. This is because the modes in resonance band are less likely to shift than when the mass depends on temperature. From the Fig.~\ref{fig:decayconstdep}, the temperature at $m=3H$ decreases as the decay constant increases. For example, if we take $f=10^{15}$ [GeV], the mass  of the axion becomes almost constant when the axion commence to oscillate. Thus, in order to investigate the possibility (iii) , it is necessary to consider the case where the decay constant of the axion is sufficiently large.  

\subsection{Nonlinear calculation} 
As discussed in the previous subsection, even if we start with an almost homogeneous initial condition, corresponding to the case where the symmetry breaking takes place before or during inflation, the tachyonic instability and/or the resonance instability can enhance the inhomogeneity to be $|\delta \phi/\phi| \sim {\cal O}(1)$. In this case, the dynamics cannot be described by the linear analysis. 
%In fact, as shown in Figs.~\ref{fig:speclinear_tac} and \ref{fig:speclinear_param}, $\delta \phi_k$ keeps on growing even after $|\delta \phi/\phi|$ reaches ${\cal O}(1)$. 
In this subsection, we solve the Klein-Gordon equation (\ref{eq:KG}) for the potential (\ref{eq:pot_align}), using (3+1) dimensional lattice simulation with $128^3$ grids. We consider three examples that correspond to type (i), (ii) and (iii), respectively, as summarized in Table~\ref{tab:IC}.
\begin{table}[htb]
\begin{center}
\caption{Simulation setup\label{tab:IC}}
\begin{tabular}{cccccc}
Type & $\langle\tilde{\phi}_i\rangle$ & $\langle\delta\tilde{\phi}^2\rangle^{1/2}$ & $\epsilon$ & $\tilde{T}_*$                & $m_{0}L$ \\ \hline
(i)      & $2\pi\times(1-10^{-9})$        & $2\pi\times10^{-8}$                        & 0.01       & 5 & $4\pi$      \\
(ii)     & $2\pi\times(1-10^{-7})$        & $2\pi\times10^{-8}$                        & 0.05        & 5 & $2\pi$      \\
(iii)    & $2\pi\times(1-10^{-2})$        & $2\pi\times10^{-8}$                        & 0.05       & 0.5 & $2\pi$     
\end{tabular}
\end{center}
\end{table}

\begin{figure}
\centering
\includegraphics[height=5.5cm,width=.48\textwidth]{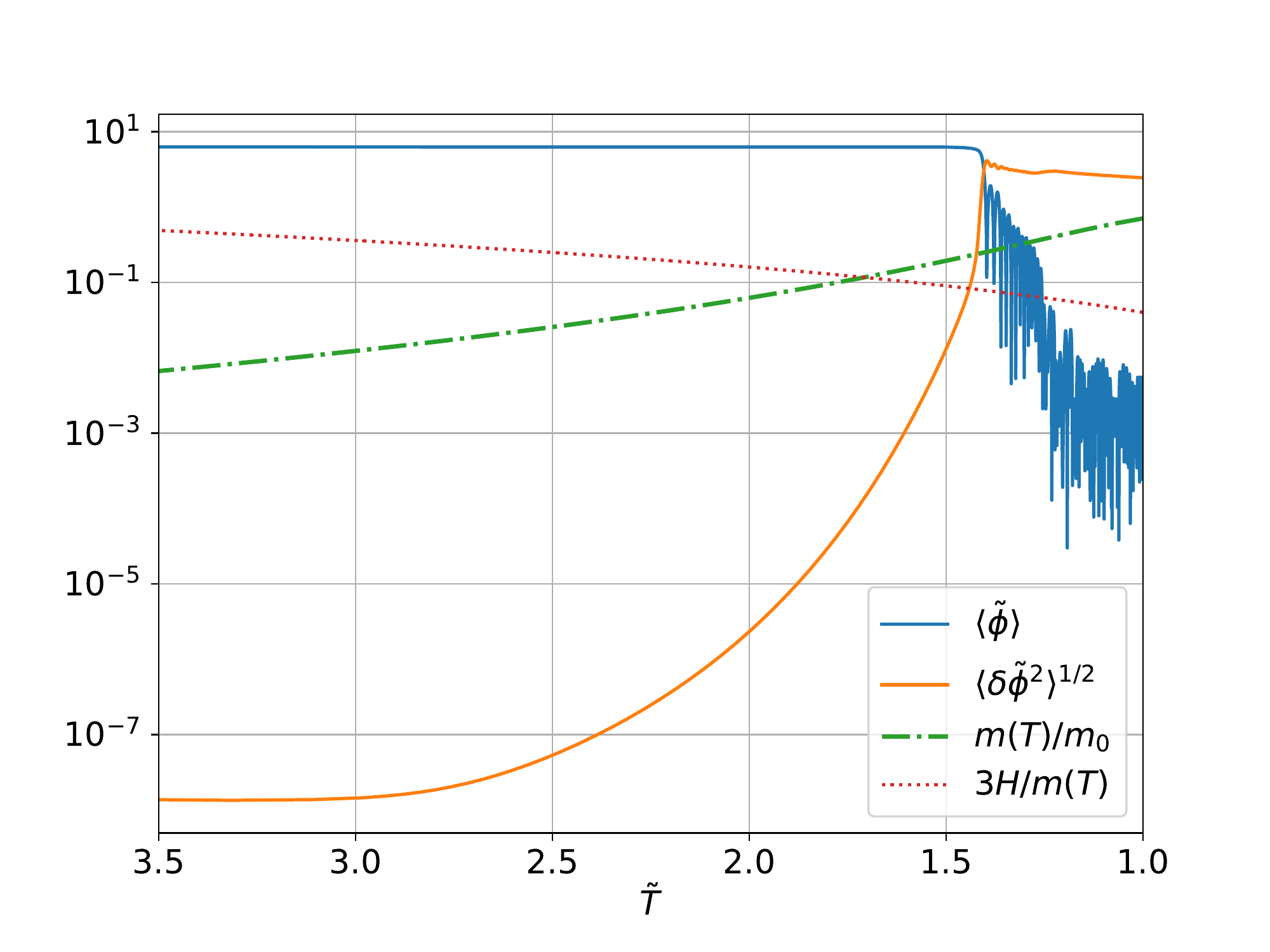}
\hfill
\includegraphics[height=5.5cm,width=.48\textwidth]{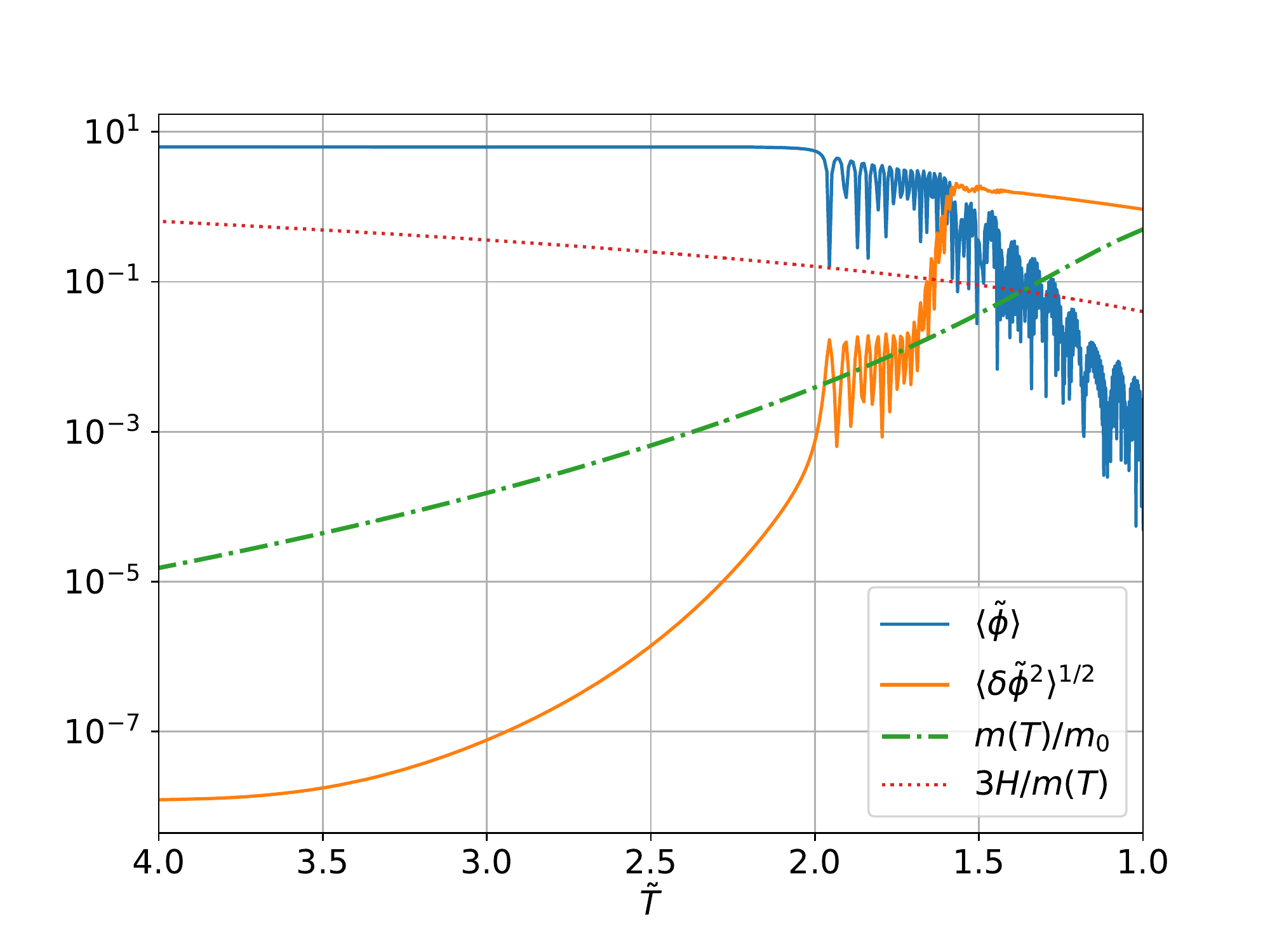}
\caption{These plots show the evolution of the spatial average of $\tilde{\phi}$(blue), the root-mean-square of field fluctuation $\langle\delta\tilde{\phi}^2\rangle^{1/2}$(orange), $m(T)/m_0$ (green dashed), and $3H/m(T)$. The left and right panels are for the type (i) and (ii), respectively.}
\label{fig:tac_tacres}
\end{figure}

\begin{figure}
\centering
\includegraphics[height=5.5cm,width=.48\textwidth]{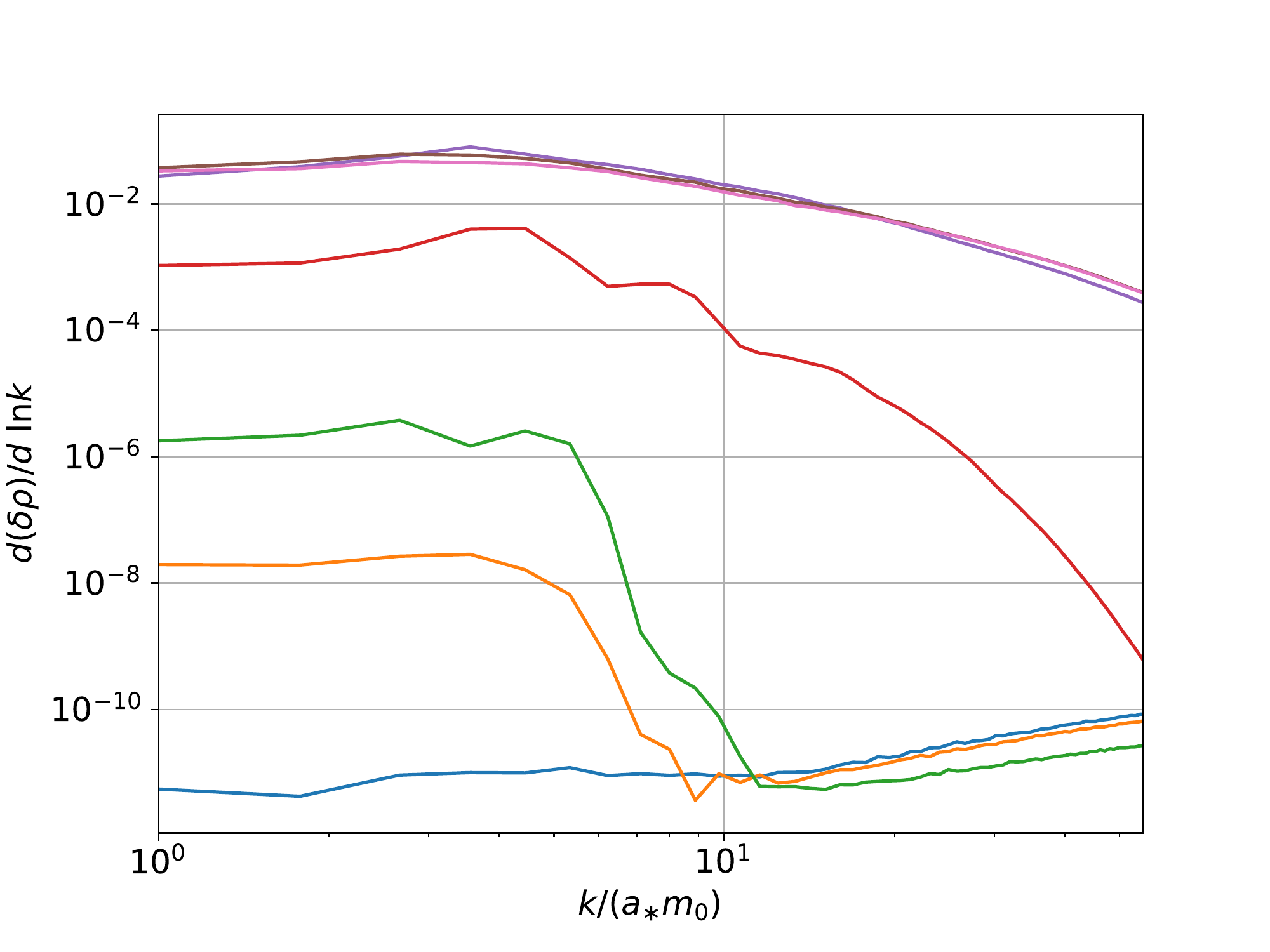}
\hfill
\includegraphics[height=5.5cm,width=.48\textwidth]{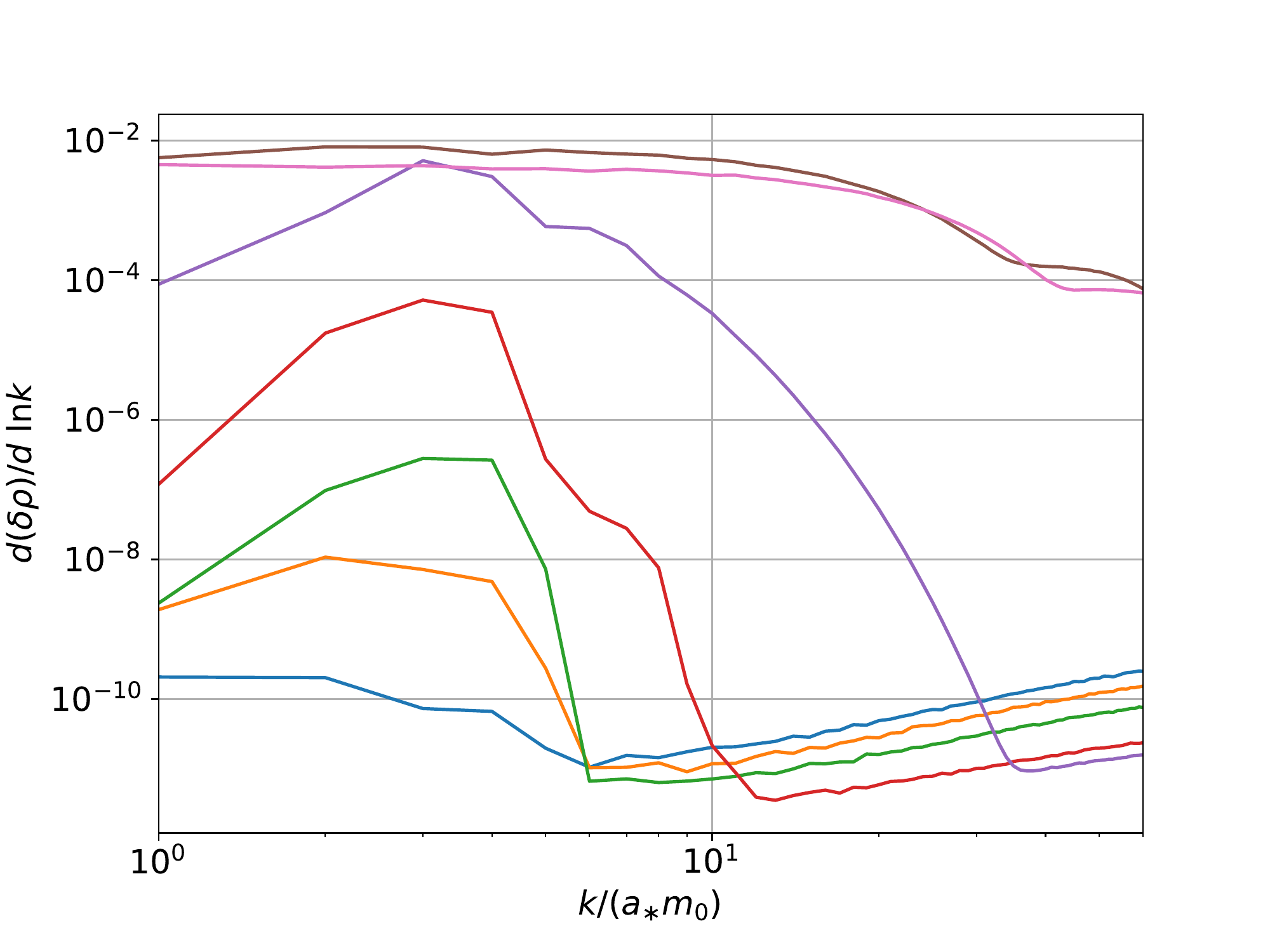}
\caption{The left and right panels show the evolution of the energy density fluctuation for type (i) and (iii). Time evolves from bottom to top. 
%{\red [NK] The label of the y-axis looks strange. It should be something like $d\rho/d\ln k$.} 
}
\label{fig:lattice_spec}
\end{figure}

\begin{figure}
    \centering
    \includegraphics[height=5.5cm]{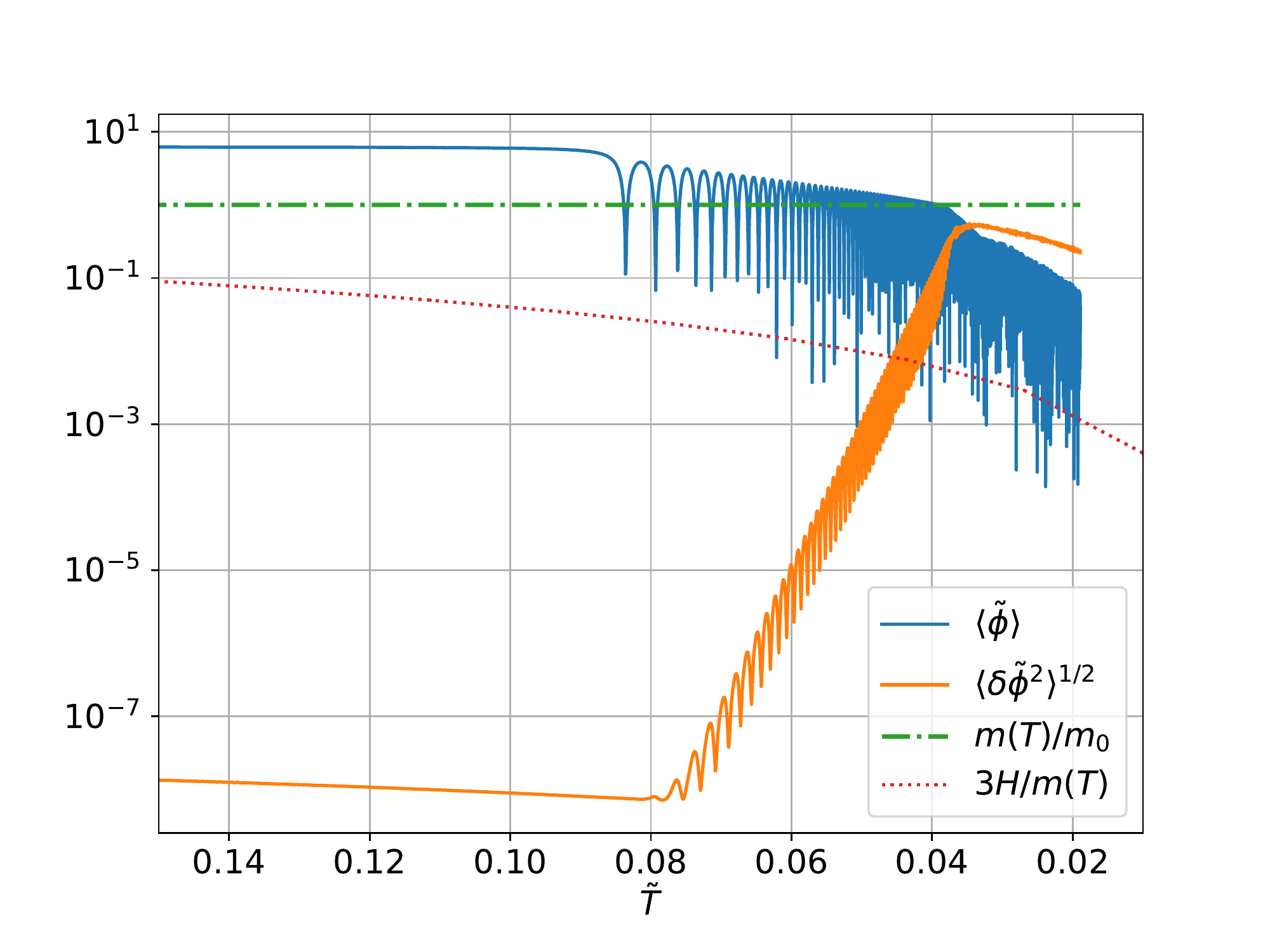}
    \caption{Same as Fig.~\ref{fig:tac_tacres} but for type (iii).}
    \label{fig:resonly}
\end{figure}

\subsubsection{Type i: Tachyonic instability}
As type (i), we consider the case where the inhomogeneity grows solely due to the tachypnic isntability. The left panel of Fig.~\ref{fig:tac_tacres} shows the evolution of the homogeneous mode (blue) and the root-mean-square of the fluctuation (orange). In type (i), the fluctuation grows exponentially due to the tachyonic instability. The green dotted line shows the time variation of the ALP mass. As the inhomogeneity grows, the spatial gradient term that disturbs the tachyonic growth becomes more prominent. When the amplitude of the inhomogeneous fluctuation becomes comparable to the homogeneous one, the tachyonic growth terminates.

\begin{figure}
    \centering
    \begin{tabular}{c}
        \begin{minipage}{0.5\hsize}
            \centering
                \includegraphics[width=5.5cm]{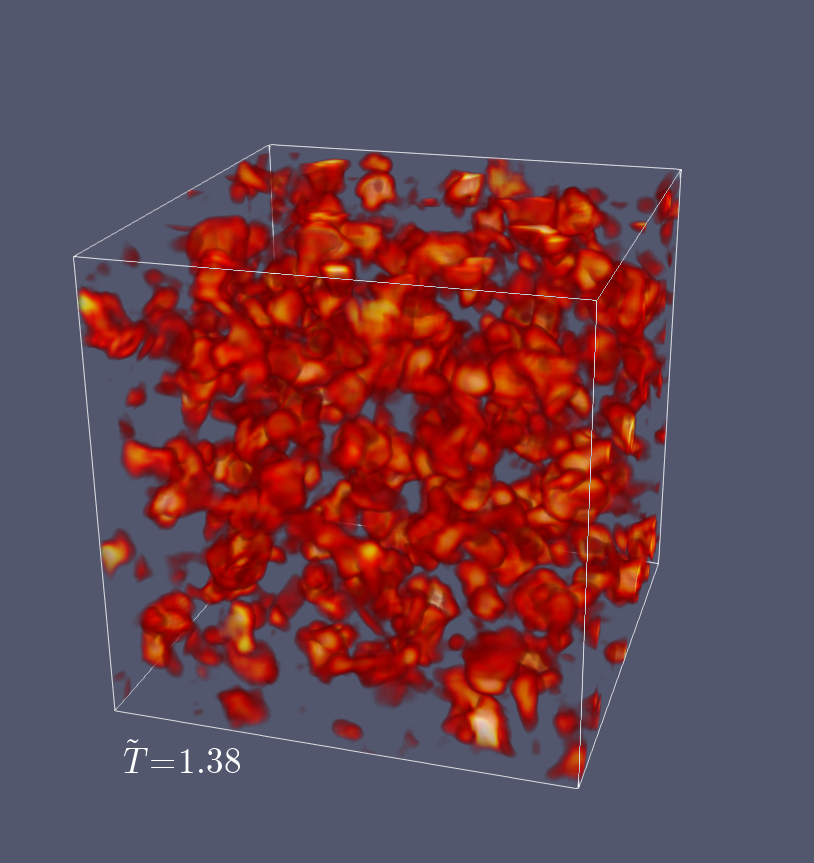}
        \end{minipage}
        \begin{minipage}{0.5\hsize}
            \centering
                \includegraphics[width=5.5cm]{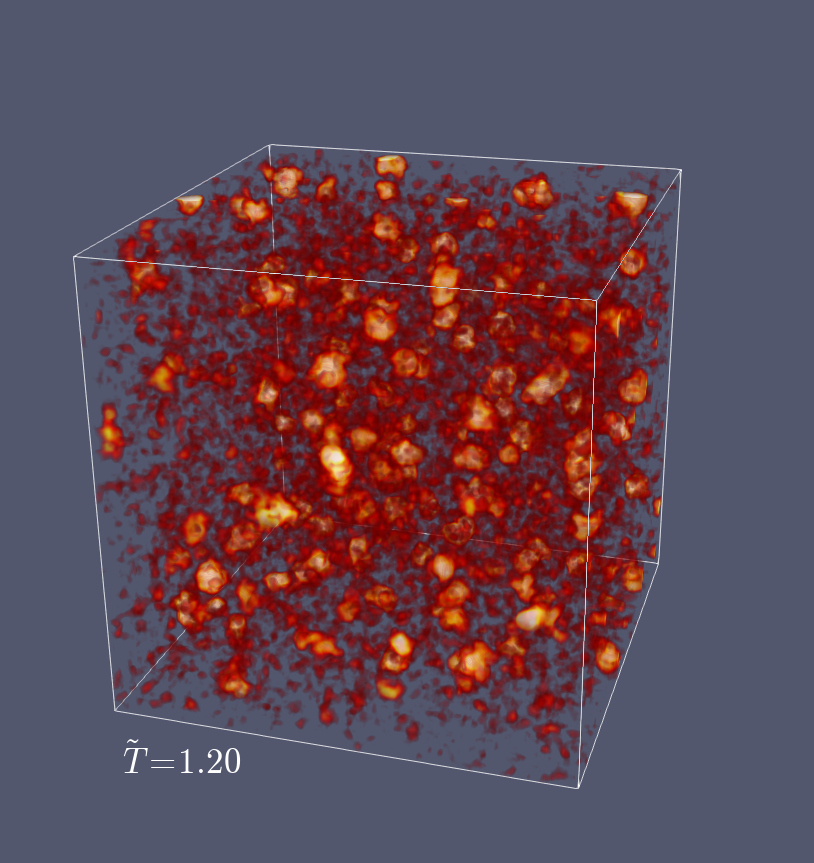}
            \end{minipage} \\
        \begin{minipage}{0.02\hsize}
        \vspace{5mm}
        \end{minipage} \\
        \begin{minipage}{0.5\hsize}
            \centering
                \includegraphics[width=5.5cm]{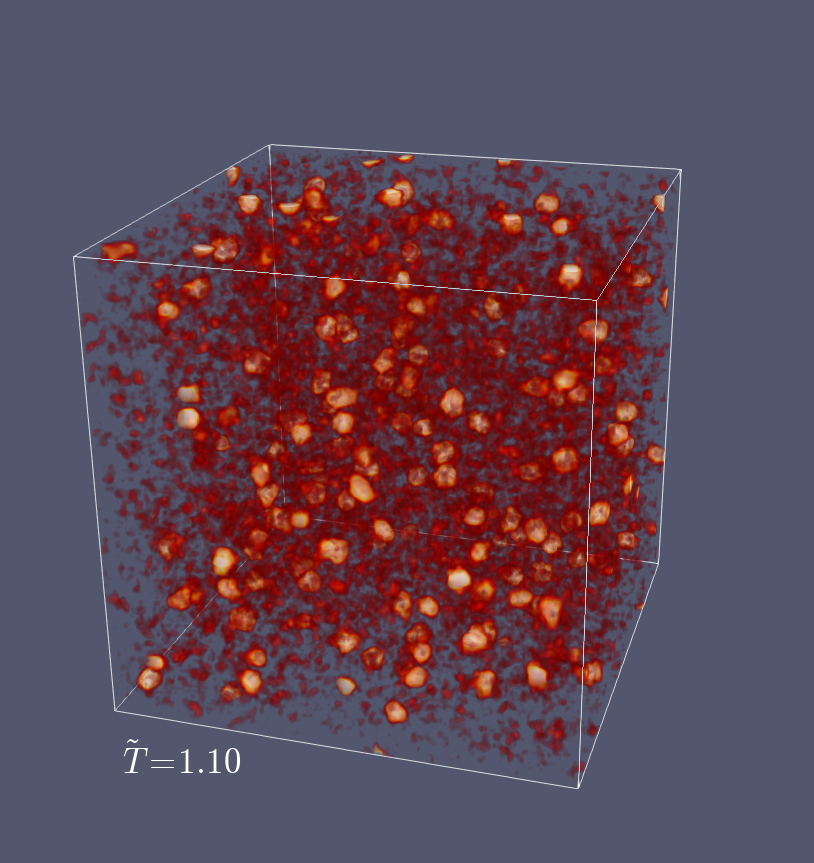}
        \end{minipage}
        \begin{minipage}{0.5\hsize}
            \centering
                \includegraphics[width=5.5cm]{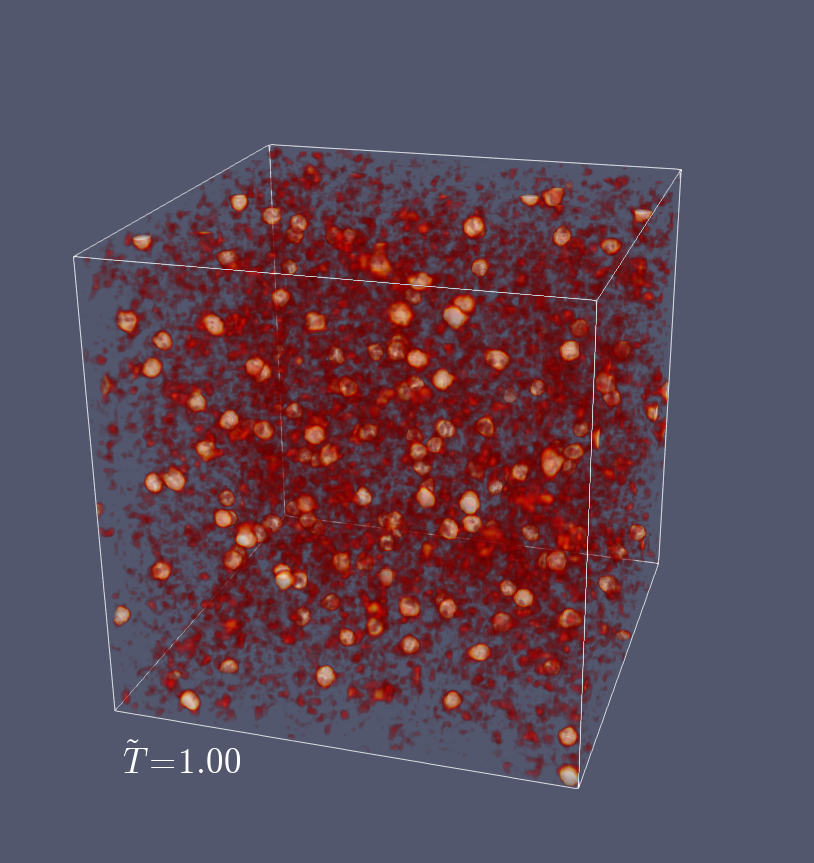}
            \end{minipage}
            
    \end{tabular}
    \caption{Snapshot of the evolution of the axion energy density evaluated at $\tilde{T}=$1.38, 1.20, 1.10 and 1.00 for type (i). The red and yellow region correspond to $\rho/\langle\rho\rangle>2$ and 4 respectively.}
    \label{fig:energyconfig_tac}
\end{figure}

The left panel of Fig.~\ref{fig:lattice_spec} shows the evolution of the spectrum of the energy density. Different colours show spectra evaluated at different times. Figure \ref{fig:energyconfig_tac} shows the snapshot of the time evolution of the energy density distribution. As is shown, the tachyonic instability leads to the formation of a number of ALP clumps, which is also known as the oscillon. Usually, oscillons are formed through the parametric resonance instability. Here, we have shown that ALP oscillons can be formed through the tachyonic instability only.

\subsubsection{Type ii: Tachyonic \& Resonance instability}
Next, we consider an example of type (ii), where the fluctuation grows due to both the tahyonic instability and the resonance instability. In this case, after the tachyonic instability stops being effective, the fluctuation starts to grow subsequently due to the resonance instability. The right panel of Fig.~\ref{fig:tac_tacres} shows the time evolution of the homogeneous and inhomogeneous modes of $\phi$. As shown by the green dotted line, the exponential growth due to the resonance instability takes place also during the ALP mass keeps on growing. This is possible because when the resonance instability started, the time scale of the oscillation has become shorter than the one of the variation of $m(T)$, which is of order of the time scale of the cosmic expansion.     

As shown in Fig.~\ref{fig:tac_tacres}, the fluctuation once stops growing for a while around $\tilde{T} \sim 2.0$. This is because the Fourier modes enhanced by the tachyonic instability differ from those enhanced by the resonance instability. The first stage of the growth is driven by the former and the second stage is driven by the latter. When the tachyonic growth terminated and the resonance instability set in, the contributions from the modes in the resonance band are still subdominant, compared to those enhanced by the tachyonic instability. Therefore, their exponential growth is not visible in the spatial average of the squared fluctuation, which sums up all the modes in the simulation. Finally, the exponential growth due to the resonance terminates, when the inhomogeneity becomes ${\cal O}(1)$ and the backreaction turns on.

\subsubsection{Type iii: Resonance instability}
For $T_*/T_c \ll 1$, when the ALP started the oscillation, $m(T)$ had already settled down at the constant value, $m_0$. The coherent oscillation with the constant period leads to the parametric resonance as discussed in the previous subsection based on the linear analysis. Figure \ref{fig:resonly} shows the time evolution of the homogeneous and inhomogeneous modes of $\phi$. Also in this case, the backreaction terminates the exponential growth due to the parametric resonance.

\begin{figure}
    \centering
    \begin{tabular}{c}
        \begin{minipage}{0.5\hsize}
            \centering
                \includegraphics[height=5.5cm]{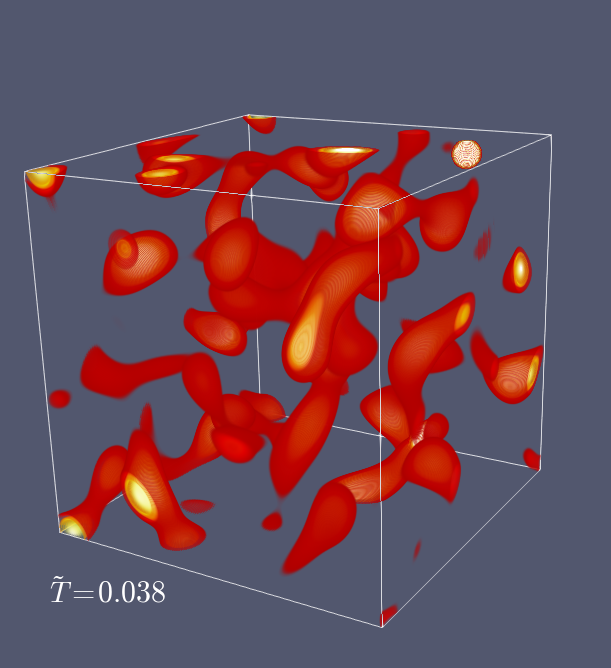}
        \end{minipage}
        \begin{minipage}{0.5\hsize}
            \centering
                \includegraphics[height=5.5cm]{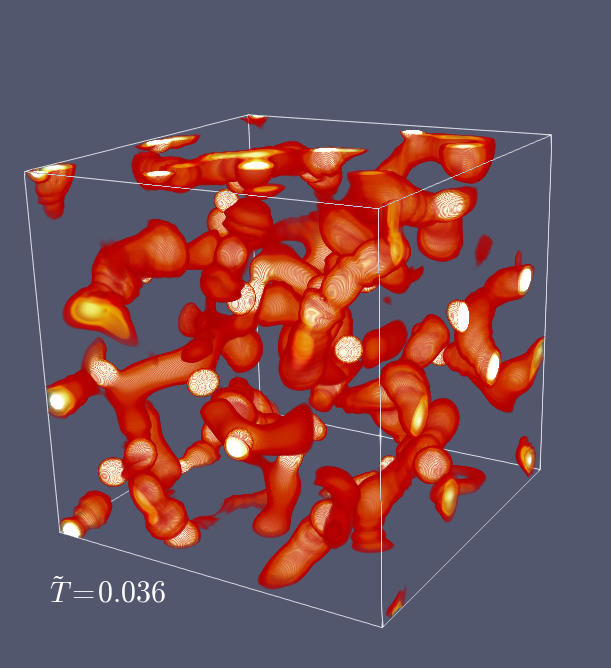}
            \end{minipage} \\
        \begin{minipage}{0.02\hsize}
        \vspace{5mm}
        \end{minipage} \\
        \begin{minipage}{0.5\hsize}
            \centering
                \includegraphics[height=5.5cm]{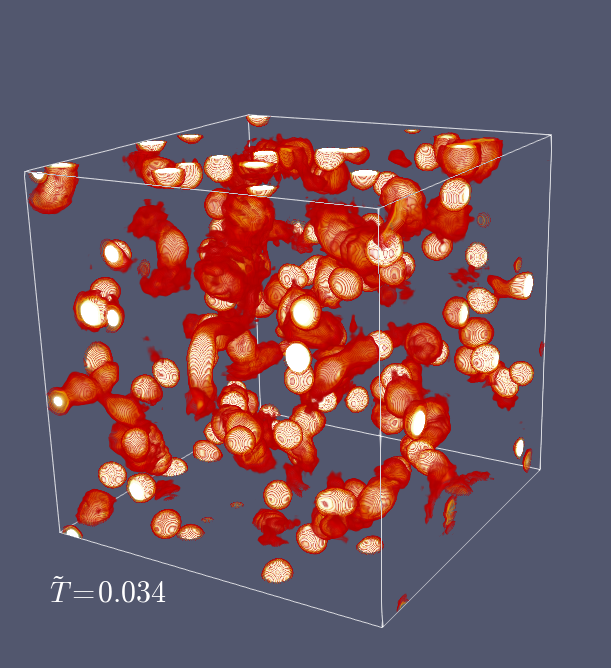}
        \end{minipage}
        \begin{minipage}{0.5\hsize}
            \centering
                \includegraphics[height=5.5cm]{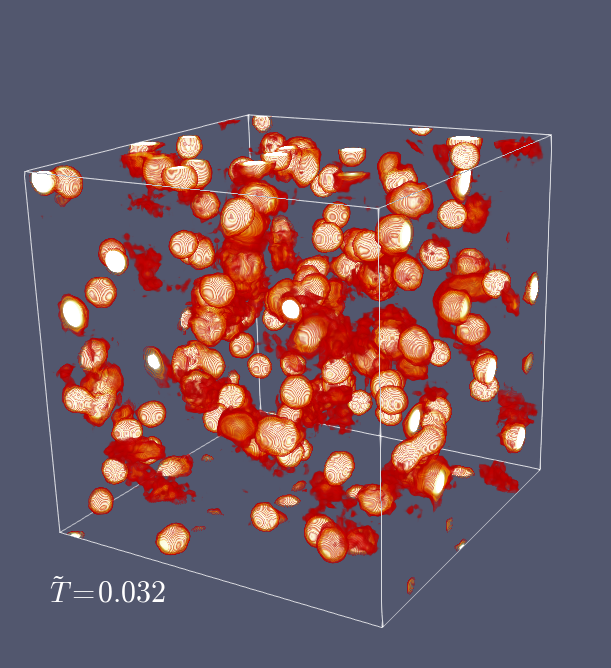}
            \end{minipage}
            
    \end{tabular}
    \caption{Snapshot of the evolution of the axion energy density evaluated at $\tilde{T}=0.038$, 0.036, 0.034, and 0.032 for type (iii). The red and yellow region correspond to $\rho/\langle\rho\rangle>2$ and 4, respectively.}
    \label{fig:energyconfig_res}
\end{figure}

The right panel of Fig.~\ref{fig:lattice_spec} shows the evolution of the spectrum of the energy density. Figure \ref{fig:energyconfig_res} shows the snapshot of the energy density. The resonance instablity also leads to the formation of a number of oscillons made of the oscillating ALP field. In Fig.~\ref{fig:energyconfig_tac}, the size of the oscillons is much smaller than the one in Fig.~\ref{fig:energyconfig_res}. This is because the typical length scale of the system $1/m$ becomes smaller and smaller due to the increasing mass for type (i).

%\begin{figure}
%\centering
%\includegraphics[height=6.3cm,width=.42\textwidth]{figure/tac_T27e-2.pdf}
%\hfill
%\includegraphics[height=6.3cm,width=.42\textwidth]{figure/tac_T20e-2.pdf}
%\caption{Snapshot of the evolution of the axion energy density for $T=, , , $, and .  Initial conditions are set for scenario (i). The red and yellow region correspond to $\rho/\langle\rho\rangle>2, 4$ respectively. }
%\label{fig:energyconfig_tac}
%\end{figure}

%\begin{figure}
%\centering
%\includegraphics[height=6.3cm,width=.42\textwidth]{figure/res_T91e-3.pdf}
%\hfill
%\includegraphics[height=6.3cm,width=.42\textwidth]{figure/res_T59e-3.pdf}
%\caption{Same as Fig.~\ref{fig:energyconfig_tac} but for $T=0.091$ [GeV] and %$T=0.059$ [GeV] in the case for scenario (iii).}
%\label{fig:energyconfig_res}
%\end{figure}

\section{Conclusion\label{sec:conc}}
In this paper, we investigated the possibility of clump formation for QCD axion and ALPs, considering the case where the PQ symmetry was already broken during inflation. For the QCD axion, while the inhomogeneity grows exponentially due to the tachyonic instability, it turned out that fluctuation does not grow sufficiently to become ${\cal O}(1)$, even for a fine-tuned initial condition. Therefore, a future discovery of the QCD axion clump strongly suggests that the PQ symmetry was broken after inflation, verifying the prevailing understanding.

Meanwhile, for ALPs, the axion clumps or the oscillons can be formed through: (i) the tachyonic instability, (ii) the tachyonic instability and the parametric resonance instability,  and (iii) the parametric resonance instability. In most of the existing studies, the oscillons are formed as a consequence of the resonance instability. In this paper, we pointed out that the oscillons can be formed through the tachyonic instability only.

In order to discuss a possible phenomenological consequence of ALP clumps or ALP oscillons, we need to understand their lifetime. According to numerical simulations, it has been known that oscillons are typically long-lived. In Ref.~\cite{Mukaida:2016hwd}, it was shown that the longevity of oscillons can be understood from an approximate U(1) symmetry, which is realized in the non-relativistic limit. The approximate symmetry correspondingly ensures an approximate conservation of particle number. Meanwhile, oscillons are not absolutely stable. As discussed in Refs.~\cite{Ibe:2019vyo, Olle:2019kbo}, the decay of oscillons is caused by a deviation from the quadratic potential, that leads to a classical emission of relativistic particles, violating the number conservation. It is interesting to see whether the lifetime of osciilons differs among type (i), (ii), and (iii). We will report this analysis in our forthcoming paper.

%During the clump formation, a prominent emission of the GWs is expected~\cite{Kitajima:2018zco}. We leave this study for our forthcoming. 

\acknowledgments
Y.~U. would like to thank G.~Moore, M.~Peloso, and D.~Schwarz for helpful discussions. We would like to thank Yukawa Institute for Theoretical Physics at Kyoto University, where a part of this work was conducted during the YITP-T-19-02 on "Resonant instabilities in cosmology". N.~K. and Y.~U. are supported by Grant-in-Aid for Scientific Research (B) under Contract No. 19H01894. N.~K. is supported by Grant-in-Aid for Scientific Research (B) 18H01243 and Grant-in-Aid for Early-Career Scientists 19K14708. Y.~U. is supported by JSPS Grant-in-Aid for Young Scientists (B) under Contract No.~16K17689, Grant-in-Aid for Scientific Research on Innovative Areas under Contract Nos.~16H01095 and 18H04349, and the Deutsche Forschungsgemeinschaft (DFG, German Research Foundation) - Project number 315477589 - TRR 211. This research was supported in part by the National Science Foundation under Grant No. NSF PHY-1748958.

\bibliography{refs.bib}

\end{document}